\documentclass[aps,pra,amsmath,amssymb,amsfonts,superscriptaddress,floatfix,nofootinbib]{revtex4} 
\pdfoutput=1
\usepackage{amsmath,amsthm,amsfonts,amssymb}
\usepackage{graphicx}

\usepackage{color}
\usepackage{hyperref}

\newtheorem{lemma}{Lemma}

\newtheorem{definition}{Definition}

\usepackage{algorithm}
\usepackage[capitalize,nameinlink]{cleveref}

\newcommand{\tM}{{\tilde M}}
\newcommand{\Gb}{G_{\rm bulk}}
\newcommand{\tS}{{\tilde S}}
\newcommand{\be}{\begin{equation}}
\newcommand{\ee}{\end{equation}}

\newcommand{\pf}{P_{\rm fail}}

\newcommand{\cG}{{\cal G}}
\newcommand{\cO}{{\cal O}}
\newcommand{\supp}{\mathrm{supp}}
\newcommand{\cS}{{\cal S}}
\newcommand{\cC}{{\cal C}}
\newcommand{\cQ}{{\cal Q}}
\newcommand{\cN}{{\cal N}}

\newcommand{\wt}{\mathrm{weight}}

\begin{document}

\title{Quantum Codes on Graphs}
\author{Matthew B.~Hastings}
\affiliation{Microsoft Quantum, Redmond, WA}
\begin{abstract}
We consider some questions related to codes constructed using various graphs, in particular focusing
on graphs which are not lattices in two or three dimensions.   We begin by considering Floquet codes which can be constructed using ``emergent fermions".  Here, we are considering codes that in some sense generalize the honeycomb code\cite{hastings2021dynamically} to more general, non-planar graphs.  We then consider a class of these codes that is related to (generalized) toric codes on $2$-complexes.  For (generalized) toric codes on $2$-complexes, the following question arises: can the distance of these codes grow faster than square-root?  We answer the question negatively, and remark on recent systolic inequalities\cite{alpert2022systolic}.
We then turn to the case that of planar codes with vacancies, or ``dead qubits", and consider the statistical mechanics of decoding in this setting.  Although we do not prove a threshold, our results should be asymptotically correct for low error probability and high degree decoding graphs (high degree taken before low error probability).
In an appendix, we discuss a toy model of vacancies in planar quantum codes, giving a phenomenological discussion of how errors occur when ``super-stabilizers" are not measured, and in a separate appendix we discuss a relation between Floquet codes and chain maps.
\end{abstract}
\maketitle

The toric code\cite{kitaev2006anyons} is a quantum stabilizer code where the qubits are arranged on a two-dimensional surface.  One can consider a variety of different lattices for the qubits, and indeed one may even consider non-translationally invariant cellulations of the surface to define the code, but in general all these forms involve some notion of geometric locality of the stabilizers.
In constrast, this paper considers quantum codes where checks are defined on general graphs, considering both Floquet codes and stabilizer codes.

In \cref{kwsec}, we give some background on Floquet codes and discuss the relation between certain Floquet codes and the Kramers-Wannier transform.  In \cref{mgfsec}, we consider Floquet codes which can be regarded as emergent Majorana fermions coupled to a $\mathbb{Z}_2$ gauge field, and discuss possible patterns of checks.
One particular class of these codes gives rise to what we call a ``(twisted) toric code on a $2$-complex", a particular kind of CSS stabilizer code; in
\cref{tc2sec} we bound the distance of this class of codes, assuming that the code is LDPC, showing that it is not possible to have
the minimum distance of the code grow faster than the square-root of the number of qubits.  As we discuss, there are examples of this code\cite{freedman2002z2,evra2022decodable} for which the \emph{product} of distances grows faster than the number of qubits, though recent systolic inequalities\cite{alpert2022systolic} show that it cannot grow \emph{polynomially} faster than the number of qubits.  However, we show that it is not possible to have both $Z$- and $X$-distances grow faster than square-root for this kind of code.  This contrasts with more general codes for which one can have distances greater than square-root\cite{hastings2021fiber} or even linear in the number of qubits\cite{panteleev2021quantum,panteleev2022asymptotically}, and these codes can in turn be used to construct linear distance toric codes on $11$-manifolds with degrees of freedom on $4$-cells\cite{freedman2021building}.

In \cref{vacsec}, we consider the case of a toric code or planar Floquet code with some vacancies or dead qubits.  We consider the case that one can measure``superstabilizers" near these vacancies; these can give rise to a check graph with high degree vertices and we consider the statistical
mechanics of decoding in this case.
Finally, in \cref{appchainmap}, we give a homological interpretation of certain Floquet codes as chain maps and in \cref{apptoy}, we give a toy model of vacancies in planar codes, describing how error probability increases with time if superstabilizers are not measured; one interesting feature is a possible superlinear growth in error probability with time.

\section{Floquet Code Background, Measuring Checks and The Kramers-Wannier Transform}
\label{kwsec}
Floquet codes\cite{hastings2021dynamically,haah2022boundaries,aasen2022adiabatic,kesselring2022anyon} are quantum codes where some operators called ``checks" are measured in some sequence, which may be periodic in time or non-periodic\cite{davydova2023floquet}.
In this paper, we consider codes on qubits, rather than qudits.
After each check is measured, in the kinds of codes considered here, the qubits will be described by some stabilizer code, stabilized by some group called the``instantaneous stabilizer group" (ISG).  This ISG is changing in time.

After measuring some check, the check itself will be part of the ISG at that time.  Any other previously measured checks $C$ will also be in the ISG if no measurement at an intervening time anti-commuted with the given check $C$.  There may also be elements of the ISG which are formed by products of checks.

All Floquet codes in this paper will involve choosing some sequence of one or two qubit measurements as checks. Each single qubit checks will measure one of the three possible Pauli operators on that qubit.  Each two qubit check will measure the product of a pair of Pauli operators; any of the $9$ possible products may be measured.

The honeycomb code and related Floquet codes (on other trivalent planar graphs with $3$-colorable plaquettes) makes use of a certain pattern of checks.  This pattern is as follows.  Consider a ring of sites (of size $6$ in the case of the honeycomb code), and label the sites by integers $1,\ldots,m$ for even $m$, with the label periodic in $m$.  Then in some round one measures a set of checks that can be chosen to be (by choosing a basis of Pauli operators on each qubit appropriately))
$X_{2j-1} X_{2j}$ for $j=0,1,\ldots,m/2$.  Then in the following round on measures a set of checks that can be
chosen to be (again by choosing a basis of Paulis)
$Z_{2j} Z_{2j+1}$ for $j=0,1,2,\ldots,m/2$.

In fact, the same pattern also occurs in the $e \leftrightarrow m$ automorphism code\cite{aasen2022adiabatic}, where this pattern is used as part of a Kramers-Wannier transform by preceding it by single qubit measurements and following with single qubit measurements.
These measurements can be chosen as follows: first measure single qubit $Z$ on all even qubits, then implement those pairwise measurements, then follow by measuring single qubit $X$ on all odd qubits.
Then this sequence implements a Kramers-Wannier transform, mapping
\be \label{KW1} Z_{2j-1} \rightarrow \pm Z_{2j-2} Z_{2j},
\ee
and mapping
\be
\label{KW2} X_{2j-1} X_{2j+1} \rightarrow \pm X_{2j},
\ee
where the signs are determined by the measurement outcomes.

Let's consider what happens in the first two measurement rounds.
Consider a given pair of qubits labeled $2j-1,2j$.  Then, we first measure $Z_{2j}$ and next measure $X_{2j-1} X_{2j}$.  The measurement $X_{2j-1} X_{2j}$ defines a stabilizer code, with a single ``logical qubit".  The effect of these two measurements is to encode qubit $2j-1$ into this logical qubit.  To see this, we can consider the two Pauli operators $X_{2j-1}$ and $Z_{2j-1}$.  The operator $X_{2j-1}$ commutes with all of these measurements and is also a logical operator of the code defined by $X_{2j-1} X_{2j}$.  The operator $Z_{2j-1}$ can be multiplied by the operator $Z_{2j}$ after the measurement of $Z_{2j}$, to give $Z_{2j-1} Z_{2j}$, which is another logical operator of the code defined by $X_{2j-1} X_{2j}$.

Similarly, the measurement of $Z_{2j} Z_{2j+1}$ defines a stabilizer code, with logical operators $Z_{2j}$ and $X_{2j} X_{2j+1}$.  After measuring $X_{2j+1}$, this effectively decodes this code, as we can multiply $X_{2j} X_{2j+1}$ by $X_{2j+1}$ to get $X_{2j}$, so these logical operators of the code defined by $Z_{2j} Z_{2j+1}$ become operator $Z_{2j}$ and $X_{2j}$ on qubit $2j$.

As shown in \cite{aasen2022adiabatic}, the effect of these four rounds of measurements (single qubit $Z$, pairwise $XX$, pairwise $ZZ$, single qubit $X$) is to implement a Kramers-Wannier transform.
So, other than these decoding and encoding operations effected by single qubit measurements, the effect of 
$Z_{2j} Z_{2j+1}$ for $j=0,1,2,\ldots,m/2$ on a state stabilized by
$X_{2j-1} X_{2j}$ for $j=0,1,\ldots,m/2$
is to map the qubits encoded in this code stabilized by $X_{2j-1} X_{2j}$ to qubits encoded in a code stabilized by $Z_{2j} Z_{2j+1}$, while performing a Kramers-Wannier transform.

This Kramers-Wannier transform effectively measures a certain product of Paulis.  Note that the map \cref{KW1,KW2} will map $\prod_j Z_{2j-1}$ to plus or minus identity.  Indeed, the measurements effectively measure this product.  Similarly, the measurements produce a state of definite $\prod_j X_{2j}$.
It should be no surprise that this happens; indeed, the product of measurements $X_{2j-1} X_{2j}$ is equal to the product of $X_j$ over all $j$, and this product commutes with the subsequent measurements of $Z_{2j} Z_{2j+1}$, and after measuring $X$ on odd qubits, we are left with definite $\prod_j X_{2j}$.

The Kramers-Wannier transform can be implemented in a different way following\cite{tantivasadakarn2021long}.
In this way, one uses single qubit $X$ measurements and controlled-$Z$ gates, instead of pairwise measurements.  The effect again is to perform the Kramers-Wannier transform while measuring a product of single qubit operators on the input state.

Unfortunately, this Kramers-Wannier transform has problems if it is used as a general technique to measure stabilizers.  That is, one might try the following.  Suppose one has some stabilizer code with some high weight stabilizer.  Without loss of generality this stabilizer can be written as a product of Pauli $X$ over some set of qubits, and so one could use those qubits as input to a Kramers-Wannier transform to measure that stabilizer, implementing that transform twice to return the qubits to their original state.
However, method of measuring stabilizers can lead to an increase in the weight of errors.  By \cref{KW2},
the map implements
$X_{2j-1} X_{2j+1} \rightarrow \pm X_{2j}$.
Hence, $X_1 X_{2j+1} \rightarrow X_2 X_4 \ldots X_{2j}$ so a two qubit on the input state can map to an arbitrarily high weight error under Kramers-Wannier transform.
Indeed, avoiding this production of high weight errors when measuring high weight stabilizers is the reason for the Shor\cite{shor1996fault}, Steane\cite{steane1997active}, and Knill\cite{knill2005quantum}
fault tolerant schemes.

\section{``Majorana-Gauge Field" Codes}
\label{mgfsec}
We now consider a class of Floquet codes where we do not use single qubit measurements and where the two qubit measurements have a particular form.  Each such code will be defined by some trivalent graph, $G$, as well as by some further data, giving a sequence of edges of the codes, which will correspond to a sequence of measurements.
There is a one-to-one correspondence between qubits of the code and vertices of the graph.

The codes we consider here have some interpretation in terms of Majorana fermions coupled to a gauge field.  They are related to codes of \cite{chapman2022free}, but we consider more general graphs.
One class of codes that we consider below is related to so-called ``matching codes"\cite{wootton2015family}.
Our codes differ (and generalize) in two ways.  First, rather than a stabilizer code with a fixed stabilizer group, we consider a Floquet code, whose stabilizer group changes in time.  Second, while this ISG at a given time is related to a matching code, it generalizes it because rather than considering a two-dimensional lattice, we consider matchings on more general trivalent graphs.

More generally, $G$ need not be a graph but may be a \emph{multigraph}.  
That is, a multigraph means that there be more than one edge between any pair of vertices; further, edges have an identity, so if there are multiple edges between a pair of vertices, different edges will correspond to different checks.
However, for brevity, we refer to $G$ as a graph.

We will assume that $G$ is connected.
The case where $G$ is not connected can be easily understood in terms of the connected case, as it will correspond to having several codes, one for each connected component.

For each of the three edges attached to a given qubit, we make an arbitrary assignment of the $3$ Pauli operators, $X,Y,Z$, choosing a different Pauli operator on each edge.  This choice has no affect on the distance or rate of the code; indeed, it has no effect on the properties of the code unless the external noise is biased in some way (for example, $X$ errors more likely than $Z$ errors).

Thus, each edge is labeled by a pair of Pauli operators, one for each of the two vertices attached to that edge.
Then each edge corresponds to a two qubit measurement, where we measure the product of the given Pauli operators on those two qubits.  For example, given an edge between vertices $1$ and $2$, where we choose Pauli $X$ for that edge on qubit $1$ and Pauli $Y$ for that edge on qubit $2$, the edge corresponds to a measurement of $X_1 Y_2$.

Then, given a sequence of edges of the code, we measure the corresponding checks in that order, repeating the sequence periodically.

There is a natural interpretation of these checks in terms of Majoranas, similar to that of the Kitaev honeycomb model\cite{kitaev2006anyons}.  For each qubit $j$, introduce four Majorana operators, $\gamma_0^j,\gamma_x^j,\gamma_y^j,\gamma_z^j$.  Impose a constraint $\gamma_0^j\gamma_x^j\gamma_y^j\gamma_z^j=1$ for all $j$.  Represent Pauli operators $X,Y,Z$ on that qubit by $i\gamma_0^j\gamma_x^j,i\gamma_0^j\gamma_y^j,i\gamma_0^j\gamma_z^j$, respectively.  See \cite{hastings2021dynamically} also.  We will not use this representation here, but it accounts for our choice of the term Majorana-Gauge Field code.
\footnote{In fact, one can consider graphs of higher degree than $3$.  In this case, if we have $d$ edges attacked to some vertex, then we take Majorana operators $\gamma_0,\gamma_1,\ldots,\gamma_d$ on that vertex.  Then, one can construct $d$ anticommuting operators on that vertex of the form $i \gamma_0 \gamma_j$ for $j\in \{1,\ldots,d\}$, with each operator having even fermion parity, and each check will be the product of one such operator on one vertex and another such operator on another vertex.  Assume $d$ odd and fix a definite value for the product $\gamma_0 \gamma_1 \ldots \gamma_d$, which commutes with all checks.  Even products of these operators $\gamma_0,\ldots,\gamma_d$ can be realized as products of Pauli operators on $m$ qubits, so long as $2^m \geq 2^{d/2-1}$.  However, for now we stick to degree $3$.}

\subsection{Homology}
These checks generate some group.  This group is sometimes called the ``gauge group" in the subsystem code literature, but we avoid using this term as it may cause confusion with other uses of the term ``gauge", e.g., ``gauge fields".  Let us define $\cG$ to be the group generated by checks, as well as by $-1$ (i.e., by a sign).  Let $\cQ$ be be $\cG$ modulo sign and let $\cN$ be the group generated by $-1$, i.e. we have a short exact sequence $1\rightarrow N \rightarrow G \rightarrow Q\rightarrow 1$.  The group $\cQ$ is abelian.

Let us determine the rank of $\cQ$.  Let there be $n_V$ vertices in graph $G$.  Hence, there are $$n_E=\frac{3}{2} n_V$$ edges.  So, $\cQ$ is generated by this set of cardinality $n_E$.  However, if we take a product of checks corresponding to some set of edges, then that product is equal (up to sign) to the identity if and only if, for each vertex, either all three edges attached to that vertex appear in the given set or no edges attached to that vertex appear in the given set.  Since we have assumed that $G$ is connected, the only nontrivial product of checks equal (up to sign) to the identity is the product of \emph{all} checks.
Hence, the rank of $\cQ$ is equal to $n_E-1$.

Let the ``stabilizer group" be the image in $\cQ$ of 
the center of $\cG$, i.e., the stabilizer group is the center of $\cG$ up to sign.  Let $\cS$ denote the stabilizer group.  We emphasize that $\cS$ is \emph{not} the ISG, though if some element of $\cS$ is in the ISG then it will remain in the ISG for all subsequent rounds.

Before considering $\cS$, we pause to define a certain homology theory.  A $1$-chain will be some formal sum of edges of $G$ with $\mathbb{Z}_2$ coefficients, i.e., we have a $\mathbb{Z}_2$ vector space with dimension equal to the number of edges, with basis elements of the vector space in one-to-one correspondence with edges.  A $0$-chain will be a formal sum of vertices of $G$ with $\mathbb{Z}2$ coefficients, and we define the obvious boundary operator $\partial$ mapping $1$-chains to $0$-chains, mapping each edge to the sum of vertices in that edge.
A $1$-cycle is a $1$-chain whose boundary vanishes.  Hence, a $1$-cycle has is a sum of edges, where for every vertex an \emph{even} number of edges are in that sum, so a $1$-cycle corresponds to a sum of closed loops on the graph, disconnected from each other.  Each such closed loop is a simple cycle, in the language of graph theory.

Corresponding to every $1$-chain is some element of $\cQ$, namely that element is the product of checks corresponding to edges whose coefficient in that chain is equal to $1$, modulo sign.
One may verify that every $1$-cycle corresponds to some product of checks which commutes with every check individually and hence corresponds to some element of $\cS$.  Let us show indeed that these $1$-cycles generate $\cS$.  Consider some edge $(i,j)$ between vertices $i$ and $j$, and consider some $1$-chain $v$.  One may verify that the product of checks corresponding to $v$ commutes with the check corresponding to $e$ if and only if $\partial v$ has the same coefficient on vertex $i$ as it does on vertex $j$.  That is, it occurs if and only if the number of edges in $v$ attached to $i$ has the same parity (mod $2$) as the number of edges attached to $j$.
However, since $G$ is assumed connected, in order for the product of checks corresponding to $v$ to commute with all checks, either $\partial v=0$ identically or $\partial v$ is equal to $1$ on every vertex, i.e., every vertex has either an even or odd number of edges attached to it in $v$.
If $\partial v=0$ identically, then $v$ is a $1$-cycle.  If $\partial v$ equals $1$ on every vertex, then $v=x+y$, where $x$ is some $1$-cycle and $y$ is the sum of \emph{all} edges.  Since the product of checks corresponding to $y$ is equal to the identity as explained above, the product of checks corresponding to $v$ is the same as the product of checks corresponding to some $1$-cycle, namely $x$.

Hence, the rank of $\cS$ is equal to the rank of the group of $1$-cycles.  This group is called the first homology group of $G$, and by the Euler characteristic of the chain complex we are studying, the rank of the first homology group is equal to $n_E-n_V+1$, since the rank of the zeroth homology group (the group of $0$-chains, modulo boundaries of $1$-chains) is equal to $1$ since $G$ is connected.
Hence, the rank of the stabilizer group $\cS$ equals $$n_E-n_V+1=\frac{1}{2}n_V+1.$$

Since $\cQ$ has rank $n_E-1=(3/2)n_V-1$, and $\cS$ has rank $s\equiv(1/2)n_V+1$, 
the difference between the rank of $\cG$ and the rank of $\cS$ is
$n_V-2$.  So, we can write $\cG$ as generated by (modulo sign)
$$\tilde Z_1,\ldots,\tilde Z_{s},\tilde X_{s+1},\tilde Z_{s+1},\ldots,\tilde X_{s+r},\tilde X_{s+r},$$
where $$r=\frac{n_V-2}{2}=\frac{1}{2}n_V-1,$$
and
where the operators $\tilde Z_i,\tilde X_j$ are products of Paulis which obey the usual Pauli anticommutation relations,
and where $$s=\frac{1}{2}n_V+1$$ is the rank of $\cS$.

After these generalities, let us consider some specific measurement sequences.  
Start with the system in a maximally mixed state, and measure checks in some sequence.  After each measurement, the state is a stabilizer state, which is stabilized by some group called the \emph{instantaneous stabilizer group (ISG)}.  Every element of the ISG is in $\cG$.

\subsection{Floquet Codes from Sequences of Perfect Matchings}
Define a \emph{perfect matching} to be a subset of edges of the graph $G$ such that every vertex is in exactly one edge in that subset.  Suppose we perform some arbitrary sequence of measurements and then measure the checks on some perfect matching.  Then, the ISG contains the checks corresponding to the edges in the perfect matching.  We claim, and we now show that \emph{the ISG is then generated by the checks in the perfect matching and possibly by some products of checks corresponding to $1$-cycles}.  Every element of the ISG commutes with these checks.  By the discussion above, if an element $O$ of the ISG (which is a product of checks, and hence corresponds to some $1$-chain $v$) commutes with a given check $(i,j)$ in the perfect matching, then $\partial v$ has the same coefficient on vertex $i$ as it does on $j$.  If $\partial v$ is equal to $1$ on $i$ and $j$, then we may multiply $O$ by the check corresponding to edge $(i,j)$; this gives some other element of the ISG, and that element corresponds to some chain $v'$ such that $\partial v'$ vanishes on $i$ and $j$.  In this way, since the matching is perfect, given any element of $O$ of the ISG, we may multiply it by checks in the perfect matching so that the resulting operator $O'$ corresponds to a $1$-cycle, as claimed.

Let $C$ be any even length simple cycle $C$ in $G$ such that half the edges in $C$ have the corresponding check in the ISG, i.e., following the edges in $C$, we see that they alternating between being in the ISG and not being in the IG, we can measure the element of $\cS$ corresponding to that cycle $C$ by measuring the checks corresponding to edges in $C$ which are not in the matching.

So, we can choose the ISG to contain edges in some matching, as well as products of edges corresponding to any cycle of even length by appropriately measuring products of checks.

So, in this subsection we consider codes for which the sequence of checks measured is given by picking some sequence of perfect matchings, called $m_1,m_2,\ldots,m_k$, for some $k$, and then measuring all the checks in $m_1$ in an arbitrary order (the order does not matter), then all the checks in $m_2$, and so on, repeating the sequence periodically with period $k$, identifying $m_{k+i}$ with $m_i$.

Given any two perfect matchings, $m,m'$, we can define a set of simple cycles that we call $C(m,m')$ as follows.
Pictorially, simply draw on the graph the edges that appear in exactly one of the two matchings (but not both of them), and this will give a set of simple cycles.
Formally, each matching $m$ or $m'$ can be identified with some $1$-chain $v_m$ or $v_{m'}$ where the chain has a coefficient $1$ on the edges in the matching.  Then, $v_m+v_{m'}$ is a $1$-cycle and hence corresponds to a union of simple cycles; these simple cycles are those in $C(m,m')$.  

It is easy to verify the following: \emph{suppose after measuring all the checks in some matching $m_i$,  the ISG is generated (up to signs) by those checks and by products of checks corresponding to $1$-cycles in some given set $S$; then after measuring all the checks in matching $m_{i+1}$, the ISG is generated by the checks in $m_{i+1}$, and by products of checks corresponding to cycles in $S$, and by products of checks corresponding to simple cycles in $C(m,m')$.}

\begin{definition}
Let us define a \emph{graph matching code} to be a stabilizer code on a trivalent graph, with ``checks" on the graph as defined above, such that the stabilizer group of the graph matching code is generated by the checks corresponding to edges of some perfect matching, and by products of checks corresponding to simple cycles, for some set of simple cycles. 
\end{definition}
Thus, what we have shown is that for such a measurement sequence using perfect matchings, after measuring any perfect matching the ISG is that of a graph matching code.

Indeed, following such a sequence of measurements, if we start from a maximally mixed state, it follows that the $1$-cycles in $S$ correspond to simple cycles, rather than sums of more than one simple cycle.  After measuring the checks in $m_{i+1}$, we may update the set $S$ to some set $S'$ containing the cycles in $S$ as well as those in $C(m,m')$.
Of course, it may be the case that there is some redundancy between these generators of the ISG, i.e., some product may be the identity.

One source of redundancy may be some redundancy in the generators which corresponds to cycles in $S'$.  If that occurs, we may remove some cycles from $S'$ to obtain a set of cycles with no redundancy.  In practice, however, this case is important for fault tolerance: such a redundancy may occur when some elements of $C(m,m')$ is already in $S$ and that is how we can detect if some error occurs because we are measuring some operator whose value is already determined by the previous measurements.

Now let us compute the number of logical qubits.  First, suppose that $S$ is maximal: every product of checks corresponding to a $1$-cycle is in the ISG.  These products of checks are simply the elements of the center of $\cG$ and the rank of the group generated by them is
$(1/2)n_V+1$ as given above.  The number of checks in the perfect matching is $(1/2) n_V$.
However, there is a redundancy: the product of all checks in the perfect matching is equal to the product of checks corresponding to some $1$-cycle.  This $1$-cycle is simply the sum of all checks \emph{not} in the perfect matching.  It is easy to verify that this is the only redundancy: suppose some product of checks in the matching times some product of checks corresponding to a $1$-cycle is equal to the identity.
Then, every vertex has either zero or three edges attached to it in this product (i.e., in the product of checks in the matching times the product of checks corresponding to $1$-cycles).  Suppose some vertex $v$ has three edges attached to in the product; then, every neighbor $w$ of $v$ also has three edges attached to it in the product, as either $w$ is attached to $v$ by a check in the matching or by a check in the $1$-cycles.  Hence, since $G$ is connected, the claim follows.
Indeed, this redundancy occurs by taking the product of all checks in the matching times the product of all checks in all the simple cycles which do \emph{not} use any checks in the matching.

So, the rank of group generated by the center of $\cG$ and by checks in a perfect matching is equal to $(1/2)n_V+1+(1/2)n_V-1=n_V$, and so if $S$ is maximal, there are no logical qubits.
If $S$ is less than maximal, there may be logical qubits; if $S$ contains all the simple cycles which do \emph{not} use any checks in the matching, then the number of logical qubits is equal $(1/2)n_V$ minus the rank of $S$.

\section{Toric Codes on $2$-Complexes}
\label{tc2sec}
Let us now consider an interesting class of graph matching codes.
We consider two types of simple cycles.  One type of simple cycle we call an ``avoiding cycle".  Such an avoiding cycle is a simple cycle that does not contain any edges of the matching.  (Deleting the edges of the matching from the graph leaves the union of all avoiding cycles.)  The other type of simple cycle we call an ``alternating cycle".  An alternating cycle is one of even length, such that the edges alternate between being in the matching and not being in the matching.

We will consider in this section graph matching codes in which the set of simple cycles $S$ defining the code includes all avoiding cycles and some set  $S'$ of other cycles which is independent of the set of avoiding cycles.  The honeycomb code is an example of such a code where $S'$ is the set of alternating cycles, but we may consider more general $S'$.  Here, by ``independent of", we regard a cycle as corresponding to a $1$-chain on the graph with $\mathbb{Z}_2$ coefficients.

We will show that, up to a low depth Clifford circuit and use of ancillas, and ignoring certain degenerate cases discussed below, these codes are equivalent to codes that we may call ``twisted toric codes on $2$-complexes".
First, note that there are two qubits in each check in the matching, but in the $+1$ eigenspace of that check there is effectively only a single qubit, i.e., a logical qubit of that code defined by that check.
Here is where the use of the low depth Clifford and ancillas comes in: we may apply a Clifford gate (such as a CNOT or other gate, depending on the check) to replace those two qubits in the check with a single effective qubit, and some ancilla.  The check forces that ancilla to be in a particular state (say $Z=+1$), and so we ignore the ancillas.

Each vertex is in exactly one avoiding cycle.  Except for certain degenerate cases, the two vertices in any given edge are in \emph{different} avoiding cycles from each other; a case where they are the same avoiding cycle is that the graph $G$ has two vertices and there are three edges connecting those two vertices.  So, ignoring the case where the cycles are the same, each effective qubit is in two checks corresponding to avoiding cycles.
We may choose (by applying single qubit Cliffords) to have these checks act as Pauli $Z$ on the effective qubit.

Each effective qubit is also in some number of checks corresponding to cycles $S'$.  These checks act as Pauli $X$ or Pauli $Y$ operators on the effective qubits in that cycle, possibly multiplied by Pauli $Z$ operator on qubits which are adjacent to that cycle, i.e., which correspond to edges attached to a vertex in that cycle.

Let us define twisted and untwisted
toric codes on a $2$-complex as follows.  We define a graph $G_{eff}$ whose vertices correspond to avoiding cycles in $G$, with an edge between two vertices if some edge in the perfect matching in $G$ connects a vertex in one avoiding cycle to a vertex in the other.
(Remark: if one really wants to include the degenerate case above where an effective qubit is in only a single avoiding cycle, this can also be treated by allowing $G_{eff}$ to have self-loops.)
We place a qubit on each edge of $G_{eff}$.  We promote $G_{eff}$ to a $2$-complex by attaching $2$-cells corresponding to cycles of $G$ in $S'$: given an cycle $C$ in $S'$, the edges in $C$ which are in the matching define a sequence of edges of $G_{eff}$.
Then, we may define the untwisted toric code on that $2$-complex by placing a qubit on each edge, placing a $Z$-stabilizer generator on each vertex, and placing an $X$-stabilizer generator on each plaquette.  We define a twisted toric code on that $2$-complex 
by placing a qubit on each edge, placing a $Z$-stabilizer generator on each vertex, and on each plaquette we place a stabilizer generator which is a product of Pauli $X$ operators on the qubits corresponding to edges in that plaquette possibly multiplied by some product of Pauli $Z$ operators supported on qubits in that plaquette or adjacent to that plaquette, and further possibly multiplied by a factor of $i=\sqrt{-1}$ if needed for Hermiticity.   

So, ignoring the degenerate case, the graph matching codes where $S$ includes all avoiding cycles are (twisted) toric codes.

In this and other $2$-complexes, we will often continue to refer to $1$-cells as edges and $0$-cells as vertices, using the terminology of graphs rather than complexes, as we will make use of some other graphs later.
 
We will say generally that a code is a (possibly twisted) toric code on a $2$-complex if some $2$-complex exists giving that code following the prescription of the above paragraph.  For us, a $2$-complex will have some graph for its $1$-skeleton, and the $2$-cells will be polygons attached to cycles of the graph.

Note indeed that a CSS code is a toric code on a $2$-complex if and only if every qubit participates in exactly two $Z$-stabilizer generators.  The ``only if" direction is obvious: each edge has two endpoints.  We will discuss the ``if" direction later, in \cref{subsecinteger}, when talking about integer homology.

The following question arises: if we consider toric codes on $2$-complexes, what is the maximum possible distance of the code?  
For an untwisted toric code,
there are in fact two distances: $d_X$, the distance against $X$ errors and $d_Z$, the distance against $Z$ errors.  The value of $d_X$ is equal to the minimum weight of a representative of a nontrivial first homology class while $d_Z$ is equal to the minimum weight of a representative of a nontrivial first cohomology class.  The weight of a representative is defined to be the number of qubits on which it acts nontrivially.

What about the distance of a twisted toric code?  We will show that, roughly speaking, the twisting obtained from graph matching codes cannot improve the distance compared to an untwisted code.  Consider some twisted toric code ${\cal T}$ that
arises from some graph matching code.  Let ${\cal T}'$ be an untwisted toric code defined in the obvious way from ${\cal T}$, using the same $Z$-stabilizer generators, and making the plaquette stabilizers products of Pauli $X$ around each plaquette.  Then, any $Z$-type logical operator $O$ of the untwisted ${\cal T}'$ is a logical operator of the twisted ${\cal T}$; one may verify that $O$ commutes with the stabilizers of ${\cal T}'$ since they are the same as those of ${\cal T}$ up to possibly multiplying by Pauli $Z$ operators and one may verify that $O$ is not in the stabilizer group of ${\cal T}$ because any nontrivial product of plaquette stabilizers of ${\cal T}$ corresponds to some nontrivial product of cycles in $S'$, and each such product (by assumption on indepdence of $S'$) will include at least one edge in the matching and hence will act as Pauli $X$ or $Y$ on at least one effective qubit.  Further, any $X$-type logical operator of the untwisted ${\cal T}$ corresponds to some cycle in $G_{eff}$ and hence to some cycle in $G$ in the obvious way (each edge in $G_{eff}$ is some edge in the matching, and we combine those edges with edges not in the matching to obtain a cycle in $G$), and that cycle in $G$ then corresponds to some logical operator of the twisted ${\cal T}$: it will be a product of Pauli $X$ on the edges in the cycle in $G_{eff}$, possibly multiplied by Pauli $Z$ on qubits in or adjacent to that cycle.

The simplest example of a toric code on a $2$-complex is to consider ``the" toric code on a cellulation of a $2$-torus.  In this case, one can obtain a code with $N$ qubits and with $d_X,d_Z$ both being $\Theta(\sqrt{N})$, where $\Theta(\cdot)$ is computer science big-O notation.  One may ask if one can do better.

If we demand that the code be LDPC (low-density parity check) also then we will show in \cref{mindb} that it is not possible that $d_X$ and $d_Z$ both be $\omega(\sqrt{N})$, where again we use big-O notation, and $\omega(\sqrt{N})$ means asymptotically larger than $\sqrt{N}$.  Here, 
\begin{definition}
A CSS code is LDPC if all $X$ and $Z$ stabilizers act on $\cO(1)$ qubits and all qubits are only in $\cO(1)$ stabilizers (note that by construction, each qubit is in exactly $2=O(1)$ $Z$-stabilizers).
\end{definition}
Remark: we use computer science big-O notation throughout.  When we give a bound on distance, later, such as saying something in $\cO(\sqrt{N})$, we mean that it is bounded by a constant times $\sqrt{N}$ with the constant depending on the constants hidden in the definition of LDPC.

\subsection{Systolic Freedom for Codes on $2$-Complexes}
There are examples of LDPC toric codes on $2$-complexes where the product $d_X d_Z$ is polylogarithmically larger than $N$.
The first example of this is in fact the first example\cite{freedman2002z2} of any code to achieve $d_X d_Z$ asymptotically larger than $N$, and it is a toric code on a cellulation of a particular $3$-manifold.  
The distances $d_X,d_Z$ are related to so-called ``systoles" of this manifold (least area surfaces representing nontrivial homology), and the property that $d_X d_Z \gg N$ is derived from a property of this manifold called systolic freedom (namely that the product of the areas of certain systoles is asymptotically larger than the volume of the manifold).  So, we use the term ``systolic freedom" to refer to $d_X d_Z \gg N$.

The second example\cite{evra2022decodable} gives $d_X=\Theta(\log(N))$ and $d_Z=\Theta(N)$, slightly improving the power of $\log(N)$ in the product $d_X d_Z$.
In this case, the code is a toric code on a particular $2$-complex which is a high dimensional expander.

To turn those toric codes with unbalanced distance (i.e., $d_Z \gg d_X$ but $d_X d_Z \gg N$) into a code with $d_Z,d_X \gg \sqrt{N}$, it is possible to use a distance balancing trick.  In the first example of a $3$-manifold code\cite{freedman2002z2}, this distance balancing was done explicitly by constructing a toric code on a \emph{four}-manifold, with degrees of freedom on \emph{two}-cells.  In later examples, an abstract distance balancing trick was used (see \cite{hastings2016weight} which was then generalized and improved in Ref.~\cite{evra2022decodable}), but when applied to the toric code on a $2$-complex, the result is again a toric code on a higher dimensional complex with degrees of freedom on cells of dimension $>1$.  So, in neither case do we obtain a toric code on a two complex.  In particular, in both cases, each qubit is in more than $2$ $Z$-stabilizers.

There is however a simple way to balance distance, albeit at the cost of breaking the LDPC property.  Choose some integer $\ell>1$, and insert $\ell-1$ vertices into each edge of the $2$-complex.  This subdivides each edge into $\ell$ edges and $\ell-1$ vertices in the obvious way.  As a result, the boundaries of $2$-cells get subdivided, and the weight of each $X$-stabilizer hence increases by a factor $\ell$, so for $\ell \gg 1$ the resulting code is not LDPC.
This subdivision multiplies the number of qubits by $\ell$, and multiplies $d_X$ by $\ell$ while leaving $d_Z$ unchanged.  So, choosing $\ell=d_Z/d_X$ we obtain a code on $N d_Z/d_X$ qubits with $d_Z=d_X$.  Applying this to the construction of Ref.~\cite{evra2022decodable}, we get a code on $N'$ qubits with $d_Z=d_X=\Omega(\sqrt{N' \log(N')})$.

Remark: the reader may verify that this subdivision of each edge into $\ell$ edges is equivalent to concatenating the given code with a repetition code with stabilizers $Z_1 Z_2, Z_2 Z_3, \ldots, Z_{\ell}$.  From standard results, this concatenation increases $d_X$ by a factor $\ell$, while leaving $d_Z$ unchanged.  The increase in weight of the $X$-stabilizers of the concatenated code then follows since we must express $X$-stabilizers in terms of logical operators of the repetition code.

Remark: one might attempt to restore the LDPC property by subdividing the $2$-cells.  Let us sketch a way to do this, giving a pictorial description.  Unfortunately, this way does not give the desired distance properties of the code, but it is worth discussing.
Consider some $2$-cell.  Its boundary is a circle, subdivided with $m=O(\ell)$ vertices.  Draw this boundary as a circle in the plane.  Draw another circle inside it, subdivided with $\lfloor m/2 \rfloor$ vertices, and draw edges from one circle to the other,
drawing the edges without crossing.
For example, label the vertices in the outer circle by $1,2,\ldots,m$ in order around the circle, periodic mod $m$, and label the vertices in the inner circle by $1',2',\ldots,(\lfloor m/2 \rfloor)'$ in order, and then draw edges from vertex $j'$ to vertices $2j,2j+1$ for each $j$.  Fill all the holes between the two circles in the drawing with $2$-cells, so that there are $2$-cells with boundary containing vertices $j',2j,2j+1$ and with boundary containing vertices $2j+1,2j+2,j',j'+1$.
Then, the local geometry is
bounded, meaning that every vertex attaches to only $\cO(1)$ edges and so that all $2$-cells between the circle have only $\cO(1)$ cells in their boundary.
Repeat this by again drawing a circle of $\approx m/4$ vertices inside the previous one, and so on, approximately halving the number of vertices each time, until finally some circle has only $\cO(1)$ vertices and can be filled with a $2$-cell.
This subdivision can be thought of as introducing some ``hyperbolic geometry" inside the outermost circle.
Unfortunately, this subdivision creates ``shortcuts" across the outermost circle, and the distance $d_X$ of the code may be reduced by a factor proportional to $\log(\ell)/\ell$.  So, it does not suffice to give us an LDPC code with distance $\omega(\sqrt{N})$.

\subsection{Systolic Almost-Rigidity}
These examples of \cite{freedman2002z2,evra2022decodable} both have $d_X d_Z$ only polylogarithmically larger than $N$.  However, there are several examples of stabilizer quantum codes where $d_X d_Z$ is polynomially larger than $N$; indeed, it is now possible to have $d_X d_Z \sim N^2$.
One may wonder: can one have $d_X d_Z$ polynomially larger than $N$ if one restricts to toric codes on $2$-complexes?

The answer is no if one restrict to LDPC codes.  This follows from a result called ``systolic almost-rigidity" for simplicial complexes\cite{alpert2022systolic}.

To use that result, first note that given any toric code into a $2$-complex, we may turn the corresponding $2$-complex into a simplicial complex (so that all $2$-cells are triangles) by triangulating each $2$-cell: pictorially, the triangulation can be seen quite simply by drawing each $2$-cell as some polygon in the plane, adding an extra vertex in the center, and drawing edges from each vertex in the polygon to the center.  This triangulation may reduce the distance $d_X$ by creating ``shortcuts" across the $2$-cell, but given that the initial code is LDPC, the distance $d_X$ is reduced by at most a constant factor.  So, it suffices to consider toric codes on simplicial $2$-complexes.

For toric codes on simplicial $2$-complexes, theorem 2 of Ref.~\cite{alpert2022systolic} shows that for any $\epsilon>0$, we have $d_X d_Z \leq \cO(N^{1+\epsilon})$.  To translate that theorem into the language of codes, their choice of a nontrivial first cohomology class $\alpha$ corresponds to the choice of a nontrivial $Z$-type logical operator.  Their quantity $\mathrm{cut}^\alpha(M)$ equals the least weight representative of that $Z$-type logical operator; see \cref{cutweight} below.  Their quantity $\mathrm{sys}_\alpha(M)$ corresponds to the minimum, over all $X$-type logical operators which anticomute with the given $Z$-type logical operator, of the least weight representative of that $X$-type logical operator.

Thus, the theorem actually says a stronger statement than that $d_X d_Z\leq \cO(N^{1+\epsilon})$.  It says that given any nontrivial $Z$-type logical operator, there is some $X$-type logical operator which anticommutes with it, and some representatives of those $Z$-type and $X$-type logical operators, such that the product of the distances is $\leq \cO(N^{1+\epsilon})$.

The authors of Ref.~\cite{alpert2022systolic} conjecture that their theorem can be strengthened to bound $d_X d_Z \leq \cO(N \mathrm{polylog}(N))$, but do not prove this.  It is also interesting to ask whether a similar result applies if we relax the LDPC assumption on the code.

Here is the lemma needed to translate their results to quantum information theory langauge.
Remark: similar results appear in \cite{alpert2022systolic} when the complex is obtained by triangulating a manifold.  There is nothing novel in our proof of the above lemma and we give it primarily to translate their results to quantum information theory terms.
\begin{lemma}
\label{cutweight}
Given a choice of $Z$-type logical operator $\alpha$, 
define a ``cut" to be a subset of qubits $H$ such that any representative of an $X$-type logical operator which anticommutes with $\alpha$ must have support on $H$.  
Define $\mathrm{cut}^\alpha(M)$ to be the minimum cardinality of a cut.   Remark: this is a paraphrase of the definition of Ref.~\cite{alpert2022systolic} in quantum information theory terms.
Then, $\mathrm{cut}^\alpha(M)$ equals the minimum weight of a representative of logical operator $\alpha$ and the minimum is attained when $H$ is the support of such a minimum weight representative.
\begin{proof}
First we show that if $O$ is a representative of $\alpha$, and if $H$ is the support of $O$, then $H$ indeed is a cut.  This is obvious, since if an operator anticommutes with logical operator $\alpha$, and hence with $O$, it must have support on the support of $O$.

Next we will show that given any cut $H$, there must be some representative $O$ of $\alpha$ supported on $H$.
Then, by the above paragraph, it follows that the minimum cut is attained on the support of a minimum weight representative.

To show that there is a representative of $\alpha$ supported on $H$, note that by CNOT gates supported on $H$, and by CNOT gates supported on the complement of $H$, we can bring the stabilizer group to the following form.  There are some stabilizers, supported on $H$ or on its complement, which are either $X$ or $Z$ on a single qubit.  
There are some qubits in $H$ which are in EPR pairs with qubits outside $H$; i.e., we have two stabilizers of the form $X X'$ and $Z Z'$ where the primed qubits are outside $H$ and the unprimed qubits are inside $H$.  Finally, there are some qubits with stabilizers $X X'$ or $Z Z'$, but not both, again where the primed qubits are outside $H$ and the unprimed qubits are inside $H$.

Now consider the logicals once the stabilizer group is in that form.
If there is a pair of qubits with stabilizer $X X'$ but not $Z Z'$, then there are logicals $X$ (or $X'$) and $Z Z'$, while if they have stabilizer $Z Z'$ but not $X X'$, then there are logicals $Z$ (or $Z'$) and $X X'$.  There may also be logical qubits where $X$ and $Z$ logical operators are both supported on $H$ or both supported on the complement of $H$.

Any $Z$-type logical operator then is of the form $O O' P Q$, where $O$ is a product of $Z$-type logical operators on some logical qubits where both $X$- and $Z$-type logicals are supported on $H$, where $O'$ is similar except the support is the complement of $H$, and where $P$ is a product of $Z$-type logical operators on logical qubits where the stabilizer group (after the transformation above) is of the form $X X'$ (i.e., $P$ can be represented by a product of operators $Z Z'$ on those qubits) and $Q$ is a product of $Z$-type logical operators on logical qubits where the stabilizer group (after the transformation above) is of the form $Z Z'$ (i.e., $Q$ can be represented by a product of operators $Z$ on those qubits).

However, by the assumption that $H$ is a cut, $O'$ must be the identity operator, as otherwise there is an $X$-type logical supported outside $H$ which anticommutes with it.  Similarly, $P$ must be the identity.
But then any $Z$-type logical operator $O Q$ has a representative on $H$.
\end{proof}
\end{lemma}

\subsection{Bound on Minimum Distance}
\label{mindb}
\begin{lemma}
\label{z2lemma}
Let $M$ be some $2$-complex corresponding to some LDPC code.  Consider the corresponding chain complex
with $\mathbb{Z}_2$ coefficients.
Let $\wt(\cdot)$ of some cycle or cocycle denote the Hamming weight.
Consider some nontrivial $1$-cocycle $S$ which has the property that $S$ cannot be written as the sum of two cocycles, each with lower weight than $S$ and such that $S$ is a minimum weight representative of some given cohomology class.
Then $\wt(S)=\cO(\sqrt{N})$ or
there is some $1$-cycle $C$ which has inner product $1$ with $S$ such that
$\wt(C)=\cO(\sqrt{N})$.
\begin{proof}
We use a graph metric for distance on the $1$-skeleton of $M$.

Let $\supp(S)$ be the support of $S$, i.e., the set of edges with nonvanishing coefficient in $S$.

Define a ``boundary set" of vertices to be the set of vertices which are in the boundary of some edge in $\supp(S)$.

Consider the following ``boundary graph" $B$: the vertex set is the boundary set and there is an edge between two vertices if the distance between them, using a graph metric on $M\setminus \supp(S)$, is bounded by the largest diameter of a $2$-cell in $M$.  We claim that this graph is connected: indeed, if not, then for any connected component, the sum of edges in $\supp(S)$ which are in the coboundary of a vertex in that connected component would define 
some cocycle with lower weight than $S$, because (by definition of the boundary graph) no $2$-cell is in the coboundary of two different edges, with one edge in the coboundary of a vertex in one connected component and the other edge in the coboundary of a vertex in another connected component.  The sum of these cocycles over connected components would equal $S$.

Consider any vertex $i$ in the boundary set.  For any $r$, let $b_r(i)$ be the set of vertices of $M$ which can be reached from $i$ by a path of length at most $r$ which avoids the support of $S$.  Here we use the graph metric on $M$ for the path length.
  
Let $c_0=\cO(1)$ be a constant chosen later.
We consider two cases: either, for some $r\leq c_0 \sqrt{N}$, there is some edge $e$ in $\supp(S)$ which is in the coboundary of two different vertices in $b_r(i)$, or there is no such edge for any such $r$.

In the first case, we can identify a $1$-cycle $C$ with weight $\leq 1+2c_0\sqrt{N}$: the edge $e$ has two vertices $j,k$ in its boundary, and there is some path $P$ of length $\leq 2c_0\sqrt{N}$ which avoids $\supp(S)$ from $j$ to $k$.  Then, concatenate $P$ with $(j,k)$ to get a path of length $\cO(\sqrt{N})$ which intersects $\supp(S)$ exactly once.  Let $C$ be the $1$-cycle which is the sum of edges in that path.

So, consider the second case.
  In an abuse of notation, let $b_r(i)$ also denote the $0$-cochain which is the sum of vertices in the given set $b_r(i)$.
Consider the $1$-cochain $S'=S+\partial^T b_r(i)$, where $\partial^T$ is the coboundary operator.
The weight of $S'$ is equal to the sum of two weights, the first being the weight of $S'$ restricted to the support of $S$, and the second being the weight of $S'$ restricted to the complement of the support of $S$.  Calls these weights $w_1,w_2$ respectively.

We claim $\wt(S)-w_1=\Omega(r)$.  Indeed,
assume $\wt(S)\geq c_0 \sqrt{N}$, as otherwise the lemma follows.
Then, since graph $B$ is connected, for $r\leq c_0 \sqrt{N}$, the intersection of $b_r(i)$ with the boundary set has cardinality $\Omega(r)$, where the constant hidden in the big-O notation depends only on the constants hidden in the definition of LDPC, not on the choice of $c_0$.  Note that by the assumption that we are in the second case, there is no edge in $S$ which is in the coboundary of two distinct vertices in $b_r(i)$.

So, since $S$ is minimal, $w_2=\Omega(r)$, as otherwise $S'$ would be lower weight than $S$.  Hence, for all $r=\cO(\sqrt{N})$, $b_r(i)$ has $\Omega(r)$ edges in its coboundary, and hence, $b_r(i)$ has cardinality $\Omega(r^2)=\Omega(c_0^2 N)$.
Let us put back in the constants hidden in the big-O notation: the cardinality of $b_r(i)$ is $\geq c' c_0^2 N$, for some constant $c'$ which depends on the constants hidden in the definition of LDPC.
However, there are only $N/2$ vertices in $M$, as there are $N$ edges.
Hence $c'c_0^2 \leq 1/2$.

Since $c'$ is fixed, we may choose $c_0$ large enough to give a contradiction, i.e., $S$ has weight $\leq c_0 \sqrt{N}$ or there is some $C$ with inner product $1$ with $S$ which has weight $\leq 2 c_0 \sqrt{N}+1$.
\end{proof}
\end{lemma}

\begin{figure}[h]
\includegraphics[width=4in]{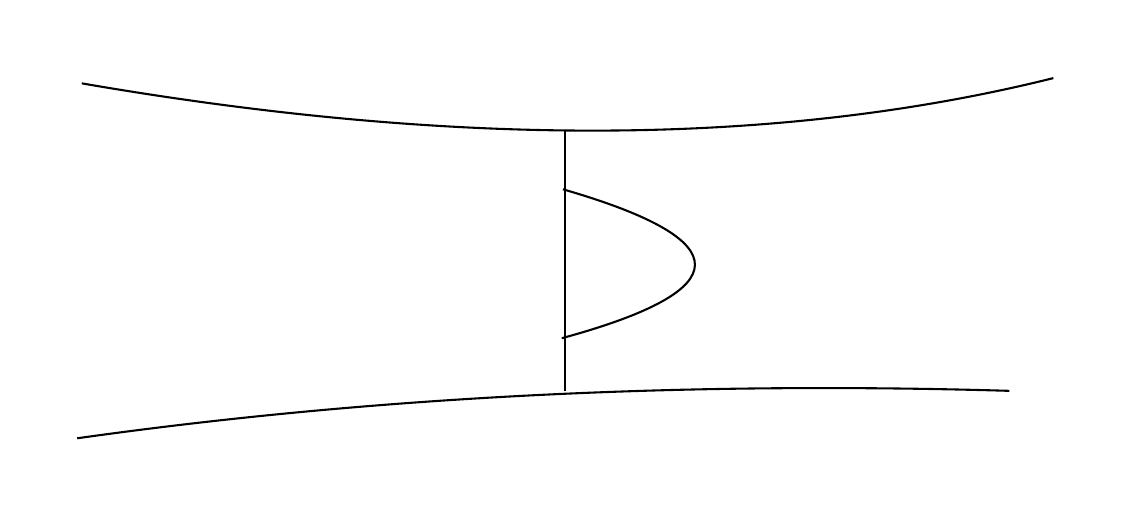}
\caption{A sketch to illustrate the proof of \cref{z2lemma}.  Suppose $M$ has the topology of a cyclinder.  The curving lines on top and bottom of the figure represent circles in the boundary of the cylinder, and the left side of the figure is attached to the right side to give a cylinder.  $M$ occupies the region between these lines; we do not show the cells in $M$ but they are implicit in the figure.  The vertical line is the support of $S$, i.e., the edges in $S$ are those which cut the vertical line.  The curving line in the middle of the figure represents part of the coboundary of $b_r(i)$, namely the part not touching the vertical line, in particular those edges in the coboundary are those cut by the curving line.  The number of edges cut by this curving line represents $w_2$.  So, as we increase $r$, the ``bubble" increases in size, and $w_2$ increases while $w_1$ decreases.}
\label{figsyst}
\end{figure}

A figure may help understand \cref{z2lemma}.  See \cref{figsyst}.

\subsection{A Remark on Distances with Integer Homology}
\label{subsecinteger}
Out of interest, let us consider some related bounds on ``distances" in the case of integer homology.  That is, we consider the homology of the chain complex \emph{with integer coefficients} associated with the given $2$-complex defined by the quantum come.  At this point, we must first remark that in the case of integer coefficients there may be more than one choice of chain complex associated with a given quantum code.  For example, suppose we have a code on three qubits, with stabilizer generators $Z_1 Z_2, Z_2 Z_3, Z_3 Z_1$ and $X_1 X_2 X_3$.  This of course corresponds to a $2$-complex with a single $2$-cell which is a triangle, with the edges corresponding to qubits $1,2,3$ respectively.
Depending on the orientation of the the triangle, one defines different chain complexes with integer coefficients which agree, mod $2$, with the $\mathbb{Z}_2$ chain complex corresponding to the given quantum code.

In \cite{freedman2021building}, such an integer chain complex was called a ``lift" of the $\mathbb{Z}_2$ chain complex.  While every $\mathbb{Z}_2$ chain complex admits a lift, in Ref.~\cite{freedman2021building}, it was conjectured that not every $\mathbb{Z}_2$ chain complex corresponding to an LDPC code admits a ``sparse lift".  Here, a ``sparse lift", means a lift such that every row and column of every boundary operator of the chain complex has the sum of the absolute values of its entries bounded by $\cO(1)$.

However, in the case of a code where every qubit participates in two $Z$-stabilizer generators, the
corresponding $\mathbb{Z}_2$ chain complex does admit some sparse lift.  This sparse lift is given by the chain complex with integer coefficients corresponding to the $2$-complex corresponding to that code.
We now explain how to construct that $2$-complex.  First define the graph corresponding to the $Z$-stabilizer generators.
Consider some $2$-cell, which corresponds to some $X$-stabilizer of form $X_1 X_2 \ldots X_k$ on some number $k$ of qubits.  Each qubit corresponds to some edge $(i,j)$ of the graph where $i,j$ are vertices.  We identify edge $(i,j)$ with edge $(j,i)$.
By the assumption that the $Z$- and $X$-stabilizers commute, the qubits $1,2,...\ldots,k$ can be ordered so that the corresponding edges define some cycle on the graph: they are in some sequence $(i_1,i_2),(i_2,_3),\ldots,(i_k,i_1)$.  Then, simply define the boundary of the $2$-cell to be those edges $(i_1,i_2),(i_2,_3),\ldots,(i_k,i_1)$ in that sequence, orienting each edge from $i_{a}$ to $i_{a+1}$.
In the corresponding chain complex, to choose a basis, we will pick an arbitrary orientation of each edge.  Thus, the boundary operator acting on the given $2$-cell will map the boundary to $\pm (i_1,i_2)+\pm (i_2,i_3)+\ldots$, where each sign depends on whether the arbitrary orientation of each edge is from $i_a$ to $i_{a+1}$ or from $i_{a+1}$ to $i_a$.

Now that we can define a chain complex with integer coefficients corresponding to some quantum code, we consider the integer homology.  A representative of homology or cohomology is some vector in a vector space with integer coefficients, and we define the weight of that representative to be the sum of absolute values of the coefficients.  We denote this weight by $\wt(\cdot)$.

\begin{lemma}
Let $M$ be some $2$-complex corresponding to some LDPC code.  Consider the corresponding chain complex with integer coefficients.
Consider some nontrivial cocycle $S$ which has the property that $S$ cannot be written as the sum of two cocycles, each with lower weight than $S$, and such that $S$ is a minimum weight representative of some given cohomology class.
Then $\wt(S)=\cO(\sqrt{N})$ or there is some $1$-cycle $C$ which has nonvanishing inner product with $S$ such that
$\wt(C)=\cO(\sqrt{N})$.
\begin{proof}
We use a graph metric for distance on the $1$-skeleton of $M$.  Let $\supp(S)$ be the support of $S$.

If necessary, we subdivide\footnote{By ``subdivide", we mean precisely that: subdivide the $1$-cell, by replacing it with $k$ $1$-cells and $k-1$ $0$-cells arranged in a line for some $k$.  This may increase the number of $1$-cells in the boundary of a $2$-cell, but this effect on the LDPC property does not matter for this proof.  This is primarily a bookkeeping device.}
the edges of $M$ so that the cohomology representative has the property that it has coefficients either  $0$ or $\pm 1$ on each edge, e.g., if $S$ has coefficient $2$ on some edge of $M$, we subdivide that edge in two and give the represenatative coefficient $1$ on each edge. 
Then, we define ``positive" and ``negative" sets of vertices: if some edge $(i,j)$ is in $\supp(\tS)$, then $i$ is in the positive set if the edge is oriented from $j$ to $i$ and the coefficient is $+1$ or if the edge is oriented from $i$ to $j$ and the coefficient is $-1$; otherwise, $i$ is in the negative set.  The vertex $j$ is in the positive set if $i$ is in the negative set and vice-versa.  If needed, we further subdivide the edges so that no vertex is in the positive set from some edge and in the negative set from some other edge, e.g., in fact if $S$ has coefficient $2$ on some edge of $M$, we should subdivide that edge in three, and give $S$ coefficients $1,0,1$ on the three edges in order; then, the first vertex is positive, the second is negative, the third is positive, and the fourth is negative.
Further, we subdivide the $2$-cells of $M$, adding additional edges, so that each two cell has only two edges with nonvanishing coefficients in its boundary.

We refer to the subdivided $M$ simply as $\tM$, and refer to the corresponding cohomology representative as $\tS$.  On the edges of $\tM$ arising from subdivision, we use any metric so long as the total length of a subdivided edge of $M$ is equal to $1$; any added edges from subdividing $2$-cells are given length $1$.

Consider the following ``positive graph" $G^+$: the vertex set is the set of positive vertices and there is an edge between two vertices if the distance between them, using a graph metric on $\tM\setminus \supp(\tS)$, 
is bounded by the largest diameter of a $2$-cell in $\tM$.  We claim that this graph is connected: indeed, if not, for each connected component, restricting $\tS$ to the edges which are in the coboundary of some vertex in that connected component would define 
some cocycle with lower weight than $S$; more precisely, it gives some cocycle of $\tM$ which pulls back to some cocycle of $M$ with lower weight than $S$.
Define a ``negative graph" $G^-$ similarly.

Let $\ell$ be the distance from the positive set to the negative set on the $1$-skeleton of $\tM$ with $\supp(\tS)$ removed, i.e., $\ell$ is the minimal length of a path from positive set to negative set avoiding $\supp(\tS)$.
We claim that $\wt(S)=\wt(\tS)\leq N/\ell$.  Indeed, consider any $d$, $0<d <\ell$.  Consider the cochain equal to $C$ plus the sum, over all vertices $v$ which can be reached by a path of distance at most $d$ which starts at the positive set and avoids the support of $\tS$, of the coboundary of $v$.
This is some other cochain which has the same inner product with $C$, and for distinct $d$ these cochains are disjoint.  By assumption that the given $S$ is minimal (and hence $\tS$ is minimal), each of these other cochains has weight $\geq \wt(S)$ and their sum is at most $N$.

Now, for any integer $k\geq 1$, define a $k$-fold cover of $\tM$ in the obvious way, using the $\tS$ as a ``cut" to define the cover\footnote{i,e, vertices in the cover are labelled by a pair, giving a vertex in the $\tM$ and an integer taken periodic mod $k$.  Going from positive to negative set in $\tS$ increases this integer by one, while going on other edges does not change the integer.}.  Call this cover $\tM(k)$.
Choose some pre-image of $\tS$ in this cover such that it gives a nontrivial cocycle of the same weight as $\tS$; call this pre-image $\tS(k)$, and define positive and negative graphs and positive and negative sets as before.   Note that $\tS(k)$ is minimal weight, as any cocycle of lower weight in the cover $\tM(k)$ would define some cocycle of lower weight than $\tS$ in $\tM$ by applying the covering map.
Now we claim that the distance from the positive to the negative set in the $1$-skeleton of $\tM(k)$ with $\supp(\tS(k))$ removed is at most $k\ell\leq k N/\wt(S)$.  Indeed, this follows by the same argument as in the above paragraph.

So, there is some path of length at most $k N/\wt(S)$ from the positive to the negative set in the $k$-fold cover, and under the covering map this defines some path in $\tM$ which starts on the positive set, goes from the negative set to the positive set $k-1$ times by an edge in $\tS$ without ever going from positive set to negative set by an edge in $\tS$, and finally ends on the negative set.  Extend this path by a single step so that it finally ends on the positive set.  Call the resulting path $P$.  This path $P$ crosses from negative to positive set a total of $k$ times by an edge in $\tS$.  Let $i_1$ be the start of path $P$.  When path $P$ crosses from negative set to positive set for the $m$-th time, call that point in the positive set $i_{m+1}$.  So, we have a sequence of points $i_1,i_2,\ldots,i_{k+1}$ on the positive set.

Pick $k=\lceil c\cdot \wt(S)/\sqrt{N}\rceil $ for a constant $c=\cO(1)$ chosen later, so that $P$ has length $\cO(\sqrt{N})$.  Note that we may assume
$\wt(S)=\Omega(\sqrt{N})$ as otherwise the lemma follows.

Since the positive set is connected, and since by assumption $\wt(S)=\Omega(\sqrt{N})$, then within distance $\sqrt{N}$ of any point in the positive set there are $\Omega(\sqrt{N})$ other points in the positive set.  Picking $c=\cO(1)$ large enough, then, by the pigeonhole principle, there must be some $m,n$ with
$1\leq m < n \leq k$ such that
some point within distance $\sqrt{N}$ of $i_m$ is also within distance $\sqrt{N}$ of $i_n$ by a path which avoids $\supp(\tS)$, and so $i_m$ is within distance
$2\sqrt{N}$ of $i_n$ by such a path; denote that path from $i_m$ to $i_n$ by $Q$.
Now form the following closed path $C$: use some segment in the middle of path $P$ to get a path from $i_m$ to $i_n$ which crosses from negative set to positive set $n-m$ times and so has inner product $n-m$ with $\tS$, and then concatenate that segment with $Q$.
This gives a closed path $C$ of total length at most $\cO(\sqrt{N})$.  This path corresponds to a $1$-cycle of weight $\cO(\sqrt{N})$ which has inner product $n-m$ with $\tS$.

If we had done some subdivision at the start of the proof, we deform the
this $1$-cycle $C$ so it lies on the $1$-skeleton of $M$, avoiding any edges in $\tilde M$ used to subdivide $2$-cells, leading to at most an $\cO(1)$ factor increase in weight.
\end{proof}
\end{lemma}
Remark: note that the inner product of $C$ with $S$ may be even.  So, if we had considered the case of $\mathbb{Z}_2$ coefficients, and we had assumed that some nontrivial cocycle $S_0$ with $\mathbb{Z}_2$ coefficients lifts to some cocycle $S$ with integer coefficients, it is possible that $C$ constructed in the above lemma might not give a cycle with $\mathbb{Z}_2$ coefficients which has nontrivial inner product with $S_0$.

\section{Vacancies and High Weight Vertices in Check Graph}
\label{vacsec}
\subsection{Vacancies and Check Graph}
An interesting question studied by several authors is how to deal with dead qubits in a planar quantum code.  Suppose qubits are chosen randomly and independently to be dead with some probability $\pf$.  For small enough $\pf>0$, for small enough noise probability $p$, is the probability of a logical error vanishing in the thermodynamic limit?

We are interested in the question of ``circuit level noise".  In a simple model of this, one alternately applies noise and measures stabilizers with some probability of error in the stabilizer measurement; the goal is to preserve logical information for some time which diverges as the system becomes large.

To have a threshold, it is necessary to measure ``super-stabilizers".  That is, dead qubits create holes in the lattice, which create extra logical qubits.  The ``super-stabilizers" measure these new logical operators.  Various schemes have been proposed in both surface and Floquet codes\cite{Tang2016,Auger2017,Nagayama2017,Strikis2021,Siegel2022,aasen2023fault} and a threshold has been proven in one particular scheme\cite{Strikis2021} with a carefully chosen schedule to measure the super-stabilizers where one has long periods (with a time depending on the size of the hole) in which the hole has one boundary condition followed by long periods in which the hole has a different boundary condition.

However, no threshold has been proven in what are perhaps the most natural schemes, in which the hole alternates its boundary conditions on a time $\cO(1)$.
These schemes lead a ``check graph" with columns of high weight vertices in spacetime, making a simple Peierls argument fail\cite{aasen2023fault}.  We analyze the statistical mechanics of this situation.
Our goal is not to prove a threshold; rather, our goal is to give give statistical mechanics arguments for the correct asymptotics of decoding in this case.

First, let us say what a check graph is.  We are considering the question of using a quantum code as a memory, where it is run for some long time, and we repeatedly measure stabilizers or checks of the code, while errors can occur at arbitrary times.  For simplicity, time is taken to be discrete.
For some codes (such as surface code or the honeycomb code), it is possible to find a basis of Pauli errors (i.e., choosing to expand errors in a particular basis of two of the three possible single qubit Pauli operators) such that
each error will cause one or two ``detection events" to occur.  If each error causes two detection events, we can construct a check graph, where each edge of this corresponds to some possible error location in spacetime and and some type (i.e. Pauli $X,Y$, or $Z$) of error, and where each vertex corresponds to some detection event, where a detection event is some particular product of measurements that should equal some fixed value $\pm 1$ if no error occurs.  An error on an edge flips the detection events in the vertices attached to that edge.  For some given error pattern (which is a $1$-chain, i.e., a sum of edges) that occurs, and some given resulting pattern of detection events, a decoder finds a decoding by choosing some other error pattern such that the sum of the two error patterns is a $1$-cycle.

We have said that an error may cause one or two detection events.  This means we also allow ``dangling edges" in a check graph.  A dangling edge only has a detection event on one of its vertices, and corresponds to an error which causes only one detection event.  These will be convenient to use later.  Edges are not dangling unless otherwise specified. 

In a code like the surface code, there actually are two distinct check graphs, one for $Z$-type errors and one for $X$-type errors.

There is a useful homological description.
We have associated error patterns with $1$-chains with $\mathbb{Z}_2$ coefficients: the chain has a coefficient $1$ on cells in the given error pattern.
 Implicitly, we also have some $2$-cells attached to the graph, any two $1$-chains in the same homology class describe the same decoding.
 
 For example, let us first describe the check graph and $2$-cells for a simple model without vacancies.
 Consider a square lattice (of size depending on the number of qubits).  There is a $2$-cell for each square in the lattice.  Call this the ``spatial check graph".  Now take this spatial check graph and cross it with a cellulation of the interval with $T$ $0$-cells for some integer $T$ giving the total time for which we run error correction, i.e., take the homological product of this $2$-complex with this $1$-complex.
 The product gives some graph in three dimensions.  Every vertex in the product then corresponds to a pair: a vertex in the spatial check graph and a time coordinate.  We describe edges in the product as ``spacelike" or ``timelike" in the obvious way: spacelike edges connect vertices with the same time coordinate.
 We use the term ``time slice" to describe a subcomplex consisting of all $0$-, $1$-, and $2$-cells at a given time coordinate.

 The spatial check graph can be considered on various topologies to give nontrivial homology, i.e., to encode logical qubits.  For example, it could be on a torus, or, to make it planar, it could correspond to a patch of surface code with ``rough" boundary conditions on one pair of opposite edges and ``smooth" boundary conditions on the opposite pair of edges.  In the second case, rough boundaries have dangling edges for one type of Pauli error, while smooth boundaries have dangling edges for the opposite type of Pauli error, and nontrivial homology representatives are described by chain stretching from one face to the opposite face.

Now we describe the check graph for these planar quantum codes with vacancies.  We actually give a ``cartoon" of the check graph, giving a slightly simplified model that skips over some details.  However,
this model captures the essential features of high degree vertices, and the results in this model should apply also to more specific models with some modifications of details.

We will call the model described here the ``vacancy model".

First, a spatial check graph is obtained in a few steps starting with the square lattice.  First, remove edges, corresponding to the dead qubits.  Remove any vertices and $2$-cells attached to removed edges.
This will create some ``holes" in the graph.  Note our terminology: a ``hole" is created by removing one or more edges corresponding to dead qubits.  At the center of each hole, add a single vertex.  We will call this vertex the ``hole center" later.
Attach one edge from the hole center to each vertex around the boundary of the hole, drawing the edges so that they do not cross, and attach $2$-cells to each region in the spatial check graph bounded by these added edges.

Importantly, note that if there are many dead qubits, the hole center may have high degree.

Now again we take this spatial check graph and cross it with a cellulation of the interval with $T$ $0$-cells for some integer $T$ giving the total time for which we run error correction, i.e., take the homological product of this $2$-complex with this $1$-complex. 

This gives some three-dimensional graph.  We will call any instead of this graph $\Gb$ below.
The ``bulk" of $\Gb$ refers to any area away from a high degree vertex.

The error model that we consider has a probability of $p$ that there is an error on each edge, independently of other edges.  This includes both spacelike and timelike edges.

Our general picture is the following.  
As is well-known and as we review below, various decoders (both Monte Carlo and minimum weight decoders) lead to error chains which are $1$-cycles by combining the actual errors with the error pattern given by the decoder.  These $1$-cycles can be decomposed into closed paths (i.e., a single cycle may decompose into several closed paths), and a logical error may occur if homologically nontrivial paths exist.
Away from the high degree hole centers, a Peierls argument (reviewed below) works to bound the probability of long chains and we in fact believe that this picture is quantitatively correct (up to subleading corrections) at small enough $p$ in the thermodynamic limit.

However, this Peiers argument breaks down near the high degree vertices.  Further, as we show below, near a high degree vertex, even a maximum likelihood decoder has its limitations.  We begin with a simplified model in which errors can occur only on edges very close to the vertex, 
and find that even a maximum likelihood decoder has a performance which is exponentially poor in the degree of the vertex.  In the simplest problem, we just consider a column of high degree vertices, with some nontrivial homology (induced by either taking it periodic in the time direction or adding dangling edges at the initial and final times), and we find that the distance in the time direction must be exponentially large in the degree to decode significantly better than random chance.
Then, of course, if a column of high degree vertices exists in a larger graph $\Gb$, the decoding can only get worse.  However, we believe that our toy model essentially captures the problem of decoding near a high degree vertex.
Our general picture then is that the dominant way in which we can have a homologically nontrivial error chain is that in the bulk the error chains are as short as possible, and can be described by a Peierls argument, but near a hole center it is hard to distinguish different error chains.  We believe that an effective model is to consider a new graph as follows.  The spatial check graph is given by taking one vertex for each hole center.  In addition, we have one vertex for every vertex in $\Gb$ with a dangling edge.
There is an edge between every pair of vertices.  Then take the product of this spatial check graph with a cellulation of an interval.  The probability of error on each edge is as follows: for a spacelike edge, the error probability is as given by a Peierls argument (roughly $p$ raised to the power of half the distance between centers), and is exponentially small in the spacing between the two hole centers (on a torus, one takes the shortest path; if homologically inequivalent paths have the same lengths, then the error probability is exponentially small in system size and so is negligible in the thermodynamic limit).  For a timelike edge, the error probability is described by our simplified models of decoding near a hole center below, and so the error probability is exponentially close to $1/2$.
Given this simplified model, it is possible that a Peierls argument works, depending on the error probabilities on the various edges: while the error probability is very close to $1/2$ on timelike edges, there is only one such path and weights for the paths to other hole centers are exponentially small.  Indeed, if the hole centers are sufficiently separated than a Peierls argument will work.  However, the Peierls argument will fail if two sufficiently high degree hole centers are sufficiently close in space.  We analyze this case in the end, and argue that a correct description is that they are replaced by a single vertex of higher degree.  For typical disorder configurations, ultimately one ends at an effective model that can be treated by a Peierls argument.

{\it An Application of this Effective Model---}
Before explaining where the effective model (with one vertex per hole center) arises, we give a brief application.  Consider a toric code on a square patch of linear size $\ell$, with smooth boundaries on two opposite sides and rough boundaries on the other sides.  If there are no vacancies, then we use a Peierls argument to estimate the error probability.  After a time $T$, the error probability is $\approx T \exp(-cL)$ for some constant $c$ depending on $p$ determined by the Peierls argument.  Of course, for large enough $T$, the error probability saturates at $1/2$..  Suppose instead we have a single hole, with hole center of degree $d$, at distance $r$ from one side and distance $\ell-r$ from the opposite side, with $r\leq \ell/2$.

Now, we can have a path of errors from one side to the opposite side going through the hole center.  For small $T$, the error probability is $\approx T^2 \exp(-cL)$.
The error probability on the timelike edge on the hole center is $1/2$ minus $\approx \exp(-c' d)$ for some some constant $c'$ depending on $p$,  Once $T$ becomes large compared to ${\rm min}(\exp(-c r), \exp(c' d))$, then the error probability crosses over to linear behavior.  If $\exp(-c r) \ll \exp(-c' d)$, then the error probability crosses over to $T \exp(c' d) \exp(-c L)$.  If $\exp(-c r) \gg \exp(-c' d)$, then the error probability crosses over to $T \exp(-c (\ell -r))$.

{\it Maximum Likelihood and Monte Carlo Decoding---}
Recall how maximum likelihood decoding works for a decoding graph. Maximum likelihood decoding can be obtained by 
defining a certain statistical mechanical model.  In this model, the only allowed error patterns are those
which give the observed syndrome, and each error pattern has a probability proportional to $(p/(1-p)))^{n_E}$, where $n_E$ is the total number of errors in the given pattern\cite{dennis2002topological}.  We can then compute the probability that the error pattern is in a given homology class, and the most likely homology class defines the maximum likelihood decoding.
This can equivalently be describing by saying:
compute, for each homology class, a ``partition function in the homology class" which is the sum over all
patterns in that class with weight $(p/(1-p))^{n_E}$, and take the homology class with the largest partition function.
The relative partition functions in different classes give the relative probabilities of the different decodings.

We can give these error patterns a pictorial description that will be useful later.
Let us color edges where an error actually occurs with a red color.  The partition function in a given homology class is a weighted sum over error patterns that give the same boundary.  Consider any such error pattern and color edges of the graph in that error pattern blue.  Thus, the edges of the graph may be colored red, blue, neither, or both.  The edges which are colored once form a closed $1$-chain.

We will refer below to the particular pattern of red colored edges as the ``quenched disorder" as that is what
it is in the context of a statistical mechanics model.

We can define a different decoder that we call a Monte Carlo (MC) decoder.  This decoder picks a random blue coloring, such that it has the same boundary as the red coloring, with a probability
proportional to $(p/(1-p)))^{n_E}$.
\begin{lemma}
Suppose that the Monte Carlo decoder has some probability $P$ of making an error in a case where there are only two homology classes.  Then, the probability that the MC decoder makes an
error is $2 P (1-P)$.
\begin{proof}
For given $P$, the probability that the correct homology class is the one found by the maximum likelihood decoder is $1-P$, while the probability that the other class is correct is $P$.
The MC decoder then picks from the class found by the maximum likelihood decoder with probability
$1-P$ while it picks from the other class with probability $P$.
\end{proof}
\end{lemma}

Note then that if $P$ is exponentially small, the probability that the MC decoder makes an error is also exponentially small.

The reason we define the MC decoder is that it is useful in results like \cref{glue} below.  We will break our analysis of the decoding problem into analyzing decoding on some subproblems, and then put those results together.  This is slightly easier to describe with the MC decoder.  The reason is, we have several probabilities to track, describing random error patterns and describing the confidence that a maximum likelihood decoder has that it gives the correct decoding on a subproblem (this confidence is important when combining problems), while for the ML decoder we can combine these into a single probability.

{\it Peierls Arguments---}
Next let us recall how a Peierls argument for a threshold works. We discuss it for both maximum likelihood and minimum weight decoders.  The Peierls argument works on graphs of bounded degree, but can break down with unbounded degree graphs.

We begin with an MC decoder.

If there are no dangling edges in the graph, then edges colored with exactly one color form a collection of closed loops, while if there are dangling edges in the decoding graph there may also be paths beginning and ending at dangling edges.  If some vertices in the decoding graph have degree greater than three, there may be some ambiguity in how to write it as loops.  For example, we could have a ``figure-eight" configuration that could be regard as either one or two loops.  This ambiguity does not matter for what is below.

For a Peierls argument, consider any path $C$, which forms either a closed loop or begins and ends at dangling edges.  Let $C$ have $|C|$ edges.  We compute the probability that every edge in $C$ is colored exactly once.  The probability that we consider here is an average over disorder (which may color some of the edges in $C$ red) of the probability (using statistical weight proportional to $(p/(1-p))^{n_E}$) that the remaining edges are colored blue and the edges colored red do not get colored blue.

Let's first emphasize what we need to calculate by giving a \emph{wrong} way of calculating this statistical weight.
To upper bound this probability, one might consider the relative statistical weight of configurations where each edge in $C$ is colored exactly once to those where no edge is colored.  There are $2^{|C|}$ possible ways of coloring each edge in $C$ exactly once.  For each way, there is a one-to-one correspondence between configurations with that coloring of edges in $C$ and configurations where no edge in $C$ is colored: simply take any configuration with the given coloring of edges in $C$, and erase the colors on all edges in $C$.  One might then naively consider the relative probability (considering again both the statistical weight of the statistical mechanical model and the probability that we there was an error pattern which gave the given red coloring of the edges in $C$) of these two configurations and think that it is $(p/(1-p))^{|C|}$.

However, this is not correct!  For any given quenched disorder, the statistical weight of the blue coloring is proportional to $(p/(1-p))^{n_E}$, but the weight depends upon the quenched disorder!
Indeed, it should be clear that the calculation is not correct: if slightly more than half of the edges in $C$ are colored red, then (taking the coloring of edges not in $C$ fixed) it is likely that the remaining edges will be colored blue, so the probability of having all edges colored once is more like $(p/(1-p))^{|C|/2}$ rather than $(p/(1-p))^{|C|}$.

Rather, a correct way is to consider each of the $2^{|C|}$ subsets of edges in $C$.  Call a subset $S$.  We consider the probability that the edges in $S$ are colored red and the edges in $C\setminus S$ are not colored red.  Then, for that $S$, we consider the probability of an event $E_S$ that the edges in $C\setminus S$ are colored blue while those in $S$ are not colored blue; to upper bound this probability, we consider the relative probability of two events of configurations, one where event $E_S$ occurs, and the other where event $E'_S$ occurs in which the edges in $S$ are colored blue while those in $ C\setminus S$ are not colored blue.  There is again a one-to-one correspondance between such configurations: simply change which edges on $C$ are colored blue.  The relative statistical weight of the two events $E_S$ and $E'_S$ is
$(p/(1-p))^{|C|-2|S|}$.
So, the probability of event $E_S$ occurring is at most
$$\frac{(p/(1-p))^{|C|-2|S|}}{(p/(1-p))^{|C|-2|S|}+1}.$$
So,
Averaging over choices of $S$, with the probability of a given $S$ being $p^{|S|} (1-p)^{|C|-|S|}$,
\begin{lemma}The probability that each edge in $C$ is colored exactly once in an MC decoder is bounded by
$$ \sum_{m=0}^{|C|} {|C| \choose m} p^{m} (1-p)^{|C|-m} \frac{(p/(1-p))^{|C|-2m}}{(p/(1-p))^{|C|-2m}+1},$$
which is exponentially small in $|C|$.
\end{lemma}

A similar Peierls argument may be made for a minimum weight decoder.  Indeed, the minimum weight decoder may be obtained by a statistical mechanical model where we sum over error patterns with some weight $(p'/(1-p'))^{n_E}$ and take a limit as $p'\rightarrow 0$.  One may use the same argument as above with this modified weight.

We can apply these Peierls arguments to prove a threshold if the maximum vertex degree is $\cO(1)$.  Consider a spatial check graph of linear size $L$ for some $L$.  Suppose all nontrivial homology representatives have length at least $L$ and end on some boundary of the spatial decoding graph at arbitrary time so that there are $\cO(L) T$ possible starting points.
If an error is made, then there must be a path of edges colored once which gives such a nontrivial representative.
For any given path of length $\ell$, the probability that the edges in that path are colored once is
$\cO(p)^\ell$, while 
the number of paths of length $\ell$ for any $\ell$ with a given start position is bounded by $\cO(1)^\ell$ where the $\cO(1)$ depends on the maximum vertex degree.
So, by a union bound, the probability of an error is bounded by
$$\sum_{\ell\geq L} \cO(p)^\ell \cO(L) T,$$ and
or small enough $p$, this is bounded by $T$ times some quantity exponentially small in $L$.

{\it Decoding With a Column of High Degree Vertices: Simplest Model---}
To begin, consider the following toy model of a high degree check graph.  We have $T$ vertices, labeled $i=0,\ldots,T-1$.  We have $d$ edges from vertex $i$ to vertex $i+1 \mod T$, for all $i$, so that each vertex has degree $2d$ and there are $Nd$ edges.

This check graph actually arises in a familiar code, a generalization of Shor's 9 qubit code.  Take a repetition code in the $Z$ basis on $T$ qubits and concatenate it with a repetition code in the $X$ basis on $d$ qubits.  Label qubits by pairs $(x,y)$ for $x\in 0,\ldots,T-1$ and $y\in 1,\ldots,d$.  Then there are stabilizers $X_{x,y} X_{x,y+1 \mod d}$ for all $x,y$, as well as stabilizers
$(\prod_y Z_{x,y}) (\prod_{z} Z_{x+1,z})$ for all $x$ for all $x$ (the $x$ coordinate is periodic in $T$ so $x=T$ is the same as $x=0$).  The check graph we have given describes correction of $X$ errors when errors occur only at a single time (i.e., we use the quantum code in a communication channel rather than as a memory).  Each vertex is a single $Z$-stabilizer.
This quantum code encodes a single logical qubit.  

Consider some random pattern of $X$ errors, choosing $X$ errors on each edge of this check graph independently with some probability $p$.
Thus, on each $d$-tuple of edges between any pair of vertices $i,i+1\mod T$, we have either an even or odd number of $X$ errors.  The probability that there are an even number of $X$ errors on that $d$-tuple equals
$$\frac{1+(1-2p)^d}{2},$$
while the probability that there are an odd number of of $X$ errors on that $d$-tuple equals
$$\frac{1-(1-2p)^d}{2},$$
so that as $d$ becomes large, at fixed $p$, both probabilities approach $1/2$ exponentially.

A decoder then finds some other pattern of errors so that in the sum of the two error patterns, all edges have the same parity of errors on each $d$-tuple, either even or odd, so that the pattern of errors found by the decoder produces the same pattern of detection events as the actual errors.  The decoder will decode correctly if the sum of the two error patterns has even parity on all edges.

Indeed, then, all that matters is whether the decoder chooses an even or odd number of errors on each $d$-tuple, so it suffices to have the decoder take either $0$ or $1$ error on each $d$-tuple, say, so that there are only two possible decodings that the decoder considers on each $d$-tuple.

This decoding problem is then the same as decoding a simpler check graph: a check graph with $T$ vertices of degree $2$, with one edge from vertex $i$ to vertex $i+1 \mod T$, for all $i$, where there is a probability $\frac{1-(1-2p)^d}{2}$ of having an error on an edge.
One may see that it requires $T$ exponentially large in $d$ for the decoder to have a decoding probability significantly better than $1/2$.
Indeed,
\emph{the probability that a maximum likelihood or MC decoder makes an error on this check graph with $T$ vertices of degree $2d$ is exponentially small in
$T (1-2p)^d$.}
To see this, note that this is the same as the problem of decoding a classical one-dimensional Ising model or repetition code: we have $T$ spins, each initialized in some unknown configuration, either all up or all down, and we flip each spin with probability $\frac{1-(1-2p)^d}{2}$.  Then, maximum likelihood decoding is the same as majority decoding, with the given probability of error.
Indeed, let's suppose that $T$ is odd.  Then error occurs if we flip more than $\lfloor T/2\rfloor$ spins, so the error probability is
$$\sum_{S\geq \lfloor T/2 \rfloor} {T\choose S} p_{eff}^S (1-p_{eff})^{T-S},$$
where $p_{eff}=(1-2p)^2$.

We may consider a slightly modified version of this check graph which has dangling edges.  Simply ``cut" the edges from vertex $0$ to vertex $T-1$, so that those two vertices each have $d$ dangling edges, and also $d$ edges going to other vertices.
Call this graph $C_{\rm column}$
Again,
\emph{the probability that a maximum likelihood or MC decoder makes an error on this check graph $C_{\rm column}$ with $T$ vertices of degree $2d$ and dangling edges is exponentially small in
$T (1-2p)^d$.}

Remark: in fact, a minimum weight matching decoder will give the \emph{same decoding} as a maximum likelihood decoder for both of these check graphs.  However, the minimum weight matching decoder is, in some sense, ``too confident" in the decoding.  That is, we can consider the minimum weight decoding for a given choice of the two possible logical decodings.
Both of
 these minimum weight decodings have either $0$ or $1$ error on each $d$-tuple of edges and they are complements of each other.  Suppose, for example, we take a fixed $T=1$; then, the difference in the weight of these two decodings is always equal to $1$, independently of $d$, even though one has only exponentially small information about the true decoding.  That is, the difference in the weights of the two different minimum weight decodings does not accurately reflect how much information we have.
Indeed, this is why we consider minimum weight and MC decoders.

Finally, it is useful to generalize this model to the case in which the edges in the check graph may have different error probabilities.
We take $T$ vertices, labeled $i=0,\ldots,T-1$.  We have $d$ edges from vertex $i$ to vertex $i+1 \mod T$, for all $i$, so that each vertex has degree $2d$ and there are $Nd$ edges.  We label the edges 
by a pair $(x,y)$ for $x\in 0,\ldots,T-1$ and $y\in 1,\ldots,d$ and we let the error probablity be $p_y$ for some $y$.  Note then that in this model the error probabilities do not depend on $x$.
Then, again
on each $d$-tuple of edges between any pair of vertices $i,i+1\mod T$, we have either an even or odd number of errors and the probability that there are an even number of errors on that $d$-tuple equals
$$\frac{1+\prod_{y=1}^d(1-2p_y)}{2},$$
while the probability that there are an odd number of of $X$ errors on that $d$-tuple equals
$$\frac{1-\prod_{y=1}^d(1-2p_y)}{2}.$$
Thus, it is the same as decoding in the case $d=1$ with an error probability on an edge equal to
$\frac{1-\prod_{y=1}^d(1-2p_y)}{2}.$
So, the probability of making an error in decoding is exponentially small in $T \prod_{y=1}^d(1-2p_y)$.

{\it Decoding Near The Hole Center---}
Now we consider a model of decoding \emph{near} a hole center of degree $d$ in the graph $\Gb$.  Specifically, we assume that all error probabilities are zero except for spacelike edges connecting some hole center to one of its neighbors and for timelike edges attached to either the hole center or one of its neighrbors.  This decoding problem is the same as taking the spatial check graph to contain just the hole center and its neighbors, with only edges from the hole center to neighbors, rather than between neighbors.

If there is no quenched disorder, then the statistical mechanics of the MC decoder here would be easy to solve using transfer matrix methods, being a one-dimensional system.  With quenched disorder, this becomes more difficult.  So, we give an approximate treatment valid for small $p$.

First, the shortest path from the center hole, back to the center hole, but at a different time, is length $3$.  This has a spacelike edge from the center hole to a neighbor at some time $s$, $0\leq s \leq T$, then a timelike edge from that neighbor to itself at a different time $s+1$, then a spacelike edge back to the center hole.  It is possible that all three of these edges are colored red.  This occurs with probability $p^3$ for any such path.  There is some ``interference" between these events: if we have such a path with some given $s$, then it overlaps with the analogous path going to the same neighbor at some time $s'=s\pm -1$.  However, for small $p$, we expect that this interference is negligible.  So, we expect that the effect is as if we considered the simplified model $C_{\rm column}$ where there are $d$ edges which each have error probability $p'=p^3$.

However, it is also possible that only some of those edges are colored red but that the decoder makes a mistake.  Indeed, we observe violated checks in the graph on some neighbor at time $s$ and at time $s+1$.  This can occur if 
 we color red the timelike edge from that neighbor at time $s$ to itself at time $s+1$, but do not color red any of the spacelike edges attached to it.  This occurs with probability $p-o(p^2)$ on any given path of three edges.  Alternatively, it can occur if we color red the two spacelike edges attached to that neighbor at times $s$ and $s+1$, which occurs with probability $p^2-o(p^3)$.  So, typically for any time $s$, there will be $\approx pd$ neighbors for which we have violated checks at the time $s$ and at $s+1$.  If we color blue the timelike edge connecting those checks, then with probability $p+o(p^2)$ we color each edge in the length $3$ path once.  So, we expect that the effect is as if we had $\approx pd$ edges
 in the simplified model $C_{\rm column}$ which each had error probability $p$.
 Again, some ``interference" is possible, between this path at some time $s$ and the analogous path at times $s'=s\pm 1$, but again for small $p$, we expect that this interference is negligible.
 
Another possibility is that we observe a violated check in the graph on some neighbor at time $s$ but no violated checks at some $s+1$ or $s-1$.
 This can occur if 
 we color red the spacelike edge from that neighbor at time $s$ to the center hole, but do not
 color red
 any of the timelike edges attached to it.
 This occurs with probability $p-o(p^2)$ on any given path of three edges.  Alternatively, it can occur if we color red the exactly one of the two timelike edges attached to it, going from that neighbor to itself at time $s\pm 1$, and also color red the spacelike edge from that neighbor at time $s\pm 1$ to the center hole.
 So, typically for any time $s$, there will be $\approx pd$ neighbors for which we have violated checks at the time $s$ but not at $s \pm1$.  If we color blue the spacelike edge from that check to the center hole,
 then w with probability $2p+o(p^2)$ we color each edge in the length $3$ path once; note the factor of $2$ in front of $2p$ due to the two possible ways in which we can make an error.  So, we expect that the effect is as if we had $\approx pd$ edges with error probability $2pd$.
 Again, some ``interference" is possible but expected to be negligible at small $p$.
 
 Thus, an approximate description should be by a model $C_{\rm column}$ with $d$ edges with effective error probability $p^3$, and $pd$ edges with effective error probability $p$ and $pd$ edges with effective error probability $2pd$.
 From the discussion above, the probability of making an error in decoding is then exponentially small in
$T \prod_{y=1}^d(1-2p_y)$, where $p_y$ are the error probabilities on these effective edges.
In this case, this is
$\prod_{y=1}^d(1-2p_y)\approx \exp(-3(p^2+o(p^3))d)$ and so the effective error
probability is exponentially small in $T\exp(-3(p^2d+o(p^3)))$.
 
Now suppose we wish to consider decoding with a single high degree hole in $\Gb$, with dangling edges from the hole center at times $0$ and $T$.  The above analysis considered only paths of length $3$ going from the hole center to itself.  We expect that longer paths are negligible, being higher order in $d$,
and so the effective error
probability will still be exponentially small in $T\exp(-3(p^2d+o(p^3)))$.

While we will not prove that the longer paths are negligible, it is possible to give a power series treatment of their effects in the limit of small $p$.  If it could be proven that this series converges, then this would give a proof of our conjecture for the decoding of a single high degree hole.  The decoder we analyze does the following.  In the first step, it does an MC decoder on a modified graph, where we remove the center hole, and replace all the edges from neighbors of the center hole to the center hole with dangling edges attached to those neighbors.  Note that the degree of this modified graph is bounded!  In the second step, it
takes the blue coloring that it finds, and uses that blue coloring on the original graph $\Gb$, coloring each edge the same way as on the modified graph; an edge from a neighbor of the hole center to the hole center is colored the same way as the dangling edge.  Having done this, there are only two possible colorings of the timelike edges on the hole center.  It picks the minimum weight such pattern.

To determine whether or not the decoder makes an error, consider each time $S$.  Consider all the timelike edges from time $S$ to time $S+1$ , including both those on the hole center and not on it.
Count the number of such edges colored red.  Also count the number of such edges colored blue in the first step.  Add these two totals, modulo $2$.  If the total is odd, then say that an ``overall error" occurs between times $S,S+1$.  If more than $T/2$ overall errors occur, then in the second step the decoder makes a logical error; if less than $T/2$ occur, then it does not.

Now, can we compute the probability that more than $T/2$ overall errors occur?  First, consider the average number of overall errors.  Pick any given time $S$.  For each timelike edge $e$ from $S$ to $S+1$, let $Z_e$ equal $-1$ if edge $e$ is colored an odd number of times, and $+1$ otherwise.
If we can compute the average of $\prod_e Z_e$, then we can straightforwardly compute the probability that there is an overall error between times $S,S+1$, as this product is $-1$ if there is an overall error.

This average is a correlation function in a statistical mechanical model with quenched disorder.  If we wanted to compute the product of the average instead of the average of the product, this would be much easier as we would be computing a product of correlation functions each involving a single degree of freedom.  This, indeed, is what happens in the graph $G_{\rm column}$: we just need to compute whether or not there are an odd number of red edges from one time to the next, and the probability that a given edge is red is independent of the other edges.

However, we can expand this correlation function (the average of $\prod_e Z_e$) as a sum of connected correlation functions.  We expect that connected correlation functions involving more degrees of freedom are suppressed in powers of $p$.  Hence, while the sum of connected correlation functions is likely not to have a well-behaved power series, we believe that the power series in $p$ for the \emph{logarithm} of the average of $\prod_e Z_e$ is a convergent expansion; the leading term of this series is the quantity $-3p^2d$ found above.

Even once we have computed the average of $\prod_e Z_e$, this is not yet sufficient, as the probability of having an overall error between times $S,S+1$ is not independent of having overall errors at other times.  However, again we believe that the power series expansion of logarithms quantities such as $\prod_e Z_e \prod_{e'} Z_{e'}$, where edges $e$ are from times $S$ to $S+1$ and edges $e'$ are from times $S'$ to $S'+1$ is a convergent expansion.

More importantly, we believe that the correlations between the overall errors at different times are negligible.
To see this, let's consider another effective model.  In this model, we consider a model of Ising spins, labelled $1,\ldots,T$, each initialized to the state $+1$, then we flip each spin independently with probability $1/2-\epsilon$ for some small $\epsilon$, and then for each pair of neighboring spins we flip both spins in that pair with probability $1/2-\epsilon'$ for some small $\epsilon'$, independently of the other pairs.  This is an effective model of the following situation: the shortest paths (of length $3$) considered above can create an overall error at some time.  However, length $4$ paths can create an overall error at two neighboring times at a higher order in $p$; we'll ignore the possibility of even further paths because they are further suppressed in $p$, but the treatment is similar to the case here.  The quantity $\epsilon$ is then exponentially small in $p^2 d$, while the quantity $\epsilon'$ is exponentially small in $p^3d$ and so $\epsilon \ll \epsilon'$.
Consider any given configuration $s_1,s_2,\ldots,s_T$ of Ising spins.  We will compute the probability that this can arise in this effective model.  This can be computed with a transfer matrix technique.  
Let $b_1,\ldots$ be binary variables corresponding to the Ising spins with $b_j=(1-s_j)/2$.
Let $f_1,f_2,\ldots$ be binary variables, such that if $f_i=1$, we flip Ising spins $i,i+1$ in the above process.

Then, the probability of any given
$b_1,b_2,\ldots$ is given by
$$c^{-1} \sum_{\{f_i\}} \prod_j z^{b_j \oplus f_{j-1} \oplus f_j} y^{f_j}$$,
where
$y=(1/2-\epsilon')/(1/2+\epsilon')$ and $z=(1/2-\epsilon)/(1/2+\epsilon)$ and where the normalization
factor $c=(1/2+\epsilon)^T (1/2+\epsilon')^{T-1}$.
Introduce matrices
$$M_0=\begin{pmatrix}
1 & z \sqrt{y} \\ \sqrt{y} z & y
\end{pmatrix},$$
and
$$M_1=\begin{pmatrix}
z & \sqrt{y} \\ \sqrt{y} & zy
\end{pmatrix},$$
so that the probability of given $b_1,b_2,\ldots$ is given by $\langle \psi_L | M | \psi_R \rangle$, where 
$$M=M_{b_2} M_{b_3} M_{b_4} \ldots M_{b_{T-1}},$$ and where the vectors $\psi_L,\psi_R$ depend on the spins $s_1,s_T$, respectively.  The exact form of the vectors $\psi_L,\psi_R$ is not so important, and we omit it (with periodic boundary conditions we replace this with a trace).
Then, it is convenient to change basis to compute the matrix product.  Introduce a new basis of vectors
$$\frac{1}{\sqrt{1+y}} (1,\sqrt{y}), \quad \frac{1}{\sqrt{1+y}}(\sqrt{y},-1).$$
This is a basis of eigenvectors of both $M_0$ and $M_1$ at $z=1$.
In this basis, we have
$$M_0=\frac{1}{1+y} \begin{pmatrix}
1+2zy+y^2 & (1-y)(1-z) \sqrt{y} \\
(1-y)(1-z)\sqrt{y} & 1-2zy+y^2
\end{pmatrix},
$$
and
$$M_1=\frac{1}{1+y} \begin{pmatrix}
z+2y+zy^2 & -(1-y)(1-z) \sqrt{y} \\
-(1-y)(1-z) \sqrt{y} & 2(z-1)y
\end{pmatrix}.
$$
Note that in this basis, for both matrices, the off diagonal terms are $\cO(\epsilon \epsilon')$, and the term in the lower right is $\cO(\epsilon+\epsilon'^2)$ for $M_0$ and $\cO(\epsilon)$ for $M_1$.  At the same time, the term in the upper left is off order unity.  Thus, so long as $T$ is small compared to
$\epsilon^{-2} \epsilon^{-2} (\epsilon+\epsilon'^2)^{-1}$, \emph{we can approximate the matrix product as the product of terms in the upper left corner}; at the same time, we will see from the solution in this regime that if $T$ is not small compared to this quantity, then the probability of a decoding error is negligible.  So, we approximate the probability as (putting in the correct normalization)
$$\frac{\Bigl(1+2zy+y^2\Bigr)^{n_\uparrow}\Bigl( z+2y+zy^2 \Bigr)^{n_\downarrow}}{\Bigl(1+2zy+y^2+z+2y+2y^2\Bigr)^T},$$ where $n_\uparrow$ and $n_{\downarrow}$ are the number of Ising spins in the $+1$ and $-1$ states, respectively.  One may recognize that this is the probability corresponding to a process where we first flip each spin independently with probability $z/(1+z)$, then flip each spin again independently with probability $y/(1+y)$, and finally again flip each spin independently with probability $y/(1+y)$.  This of course is correct for any given spin in the original effective model: each spin is subject to those three possible spin flips, however the important thing is that we can ignore any correlation between flips of neighboring spins.

{\it Gluing Together---}
Next we have a general principle.  We have discussed decoding in various cases with dangling edge, either considering decoding near a hole with dangling edges at some time, or considering Peierls arguments for some path with dangling edges.  Now suppose we have two such decoding problems.

\begin{lemma}
\label{glue}
Consider two separate decoding graphs $G_1,G_2$, each with dangling edges.  Suppose the probability that an MC decoder makes an error on graph $G_1$ is at most some probability $P_1$, and similarly the probability that it makes on error on graph $G_2$ is at most some $P_2$.
Consider the graph obtained by attaching some dangling edge $e_1$ of $G_1$ to some other dangling edge $e_2$ of $G_2$, and assume there are only two homology classes on $G_1$, and on $G_2$, and on the combined graph, so that making any error on the combined graph requires coloring $e_1$ and $e_2$ both once.
.  The probability that it makes an error on that graph is at most
$2P_1 P_2$.
\begin{proof}
Consider some fixed quenched disorder on each graph.  Suppose for that given quenched disorder, the
MC decoder makes an error with probability $Q_1$ on the first graph and probability $Q_2$ on the second graph.  The MC decoding is two separate decoding problems, constrained by the requirement that $e_1$ must be colored the same, mod $2$, as $e_2$.
So, the probability of an error on the combined graph is
$$\frac{Q_1 Q_2}{Q_1 Q_2+(1-Q_1)(1-Q_2)}\leq 2 Q_1 Q_2.$$

Further, we know that the average of $Q_1$ over quenched disorder is equal to $P_1$ and the average of $Q_2$ is equals to $P_2$.  So, the average of $2Q_1 Q_2$ equals $2 P_1 P_2$.

Remark: this worst case is achieved if there is a probability $2 P_1$ that the quenched disorder is such that $Q_1=1/2$ and similarly a probability $2P_2$ that $Q_2=1/2$.
\end{proof}
\end{lemma}

Then, in our general picture at the start, we believe that we can use the Peierls argument for paths between columns of high degree vertices, use the effective model $C_{\rm column}$ above for high degree vertices with dangling edges on the vertices where a Peierls path arrives at a vertex, and use this lemma to glue the paths, to get the effective with weak edges connecting different hole centers discussed at the start.

{\it Two Nearby High Degree Vertices---}
Finally, suppose we have two nearby high degree vertices in our effective model with one vertex per hole center and edges between every pair of vertices.  We can describe these two nearby high degree vertices by a check graph with a ``ladder" shape.  The ``rungs" of the ladder are the spacelike edges between vertices.  Suppose we have some error probability $r$ on this edge, while we have error probabilities $p_1=1/2-\epsilon_1$ and $p_2=1/2-\epsilon_2$ on timelike edges attached to one or the other hole center.

If $r\ll \epsilon_1^{-1}$ or $r \ll \epsilon_2^{-1}$, then we can use a Peierls argument on this effective ladder model.  However, suppose $r \gg \epsilon_1$ and $r \gg \epsilon_2$.  Then, a Peierls argument does not work.  In this case, in the limit of large $r\epsilon_1^{-1}$ and large $r \epsilon_2^{-1}$, we cannot effectively determine whether or not any of the spacelike edges are colored red.  Then, we arrive at an effective model $C_{\rm column}$ with $d=2$ and with error probabilities $p_1=1/2-\epsilon_1$ and $p_2=1/2-\epsilon_2$.
Thus, this is the same as a model $C_{\rm column}$ with $d=1$ and error probability
$$\frac{1-\prod_{y=1}^2(1-2p_y)}{2}=1/2-2\epsilon_1 \epsilon_2.$$

\appendix
\section{Chain Maps}
\label{appchainmap}
Many constructions in homological algebra have some relationship to some construction in quantum error correcting codes.  Most obviously, quantum codes correspond to chain complexes over $\mathbb{F}_2$, and logical $X$ and $Z$ operators correspond to homology or cohomology classes.

Here we will discuss an interesting relationship between measurements on quantum codes and chain maps.

Consider the case of a CSS quantum code $C$.  We may define a corresponding chain complex $\cC$ with three degrees, which we label by $Z,X,Q$, with $Z\rightarrow X \rightarrow Q$, and two boundary operators, $\partial_Z:Z\rightarrow Q$ and $\partial_Q:Q\rightarrow X$.  The different degrees are each given some preferred basis and $Z$-stabilizer generators correspond to basis elements of $Z$, qubits to basis elements of $Q$, and $X$-stabilizer generators to basis elements of $X$, with the boundary operator determined by the stabilizers of the quantum code.
See Ref.~\cite{bravyi2014homological} for a review and a dictionary between these languages.

Remark: this reverses the order we used previously, where we used $Z$-stabilizers for $0$-cells and $X$-stabilizers for $2$-cells.  It is chosen for consistency with other literature.

While this mapping to stabilizers seems restricted to the case of CSS codes, given any more general quantum stabilizer code on qubits, we may construct a corresponding CSS code which is self-dual.  See Ref.~\cite{bravyi2010majorana} for the general construction.  Roughly, this is done by regarding such a more general code as an instance of a code whose stabilizers are products of Majorana operators, encoding each qubit into $4$ Majoranas; then, the stabilizers of this Majorana code can be interpreted as defining both the $X$- and $Z$-stabilizers of some self-dual CSS code.

So, we will consider just this case of CSS codes.

Now, suppose we measure some operator $O$ which is a product of Pauli $X$ and is not in the stabilizer group; the case of measuring a product of Pauli $X$ is similar.  Suppose $O$ does not commute with one of the $Z$-stabilizers (so $O$ is not a logical operator of the code).  As a result, this measurement of $O$ does not increase the rank of the stabilizer group: it increases the rank of the group generated by $X$-stabilizers by one but it decreases the rank of the group generated by $Z$-stabilizers by one.  Of course, in the case of a self-dual code derived as above, a measurement of an arbitrary operator would map to a measurement of some product of Pauli $X$ and ``the same" product of Pauli $Z$; since these two measurements commute (necessarily it is an even weight product) we could do them in an arbitrary order.

We identify this operator $O$ with some vector $w\in Q$.
This vector is a sum of basis elements corresponding to qubits in the product defining $O$.
By assumption, there is some $Z$-stabilizer which anticommutes with $O$; choose an arbitrary such $Z$ stabilizer and identify that $Z$-stabilizer with some vector $v\in Q$, that vector being a sum of basis elements in the product defining the given $Z$-stabilizer.
Then,
the anticommutation can be expressed as
$$\langle v,w \rangle=1,$$
where the angle brackets denote an inner product.

After this measurement, we have some new stabilizers of some new code $C'$ and some corresponding chain complex $\cC'$, $Z'\rightarrow X' \rightarrow Q'$, with boundary operators $\partial'_{Z'}$ and $\partial'_{Q'}$.  Of course, $Q'$ and $Q$ are of the same dimension.

We will define a chain map $f$ from $\cC$ to $\cC'$.  A chain map is a collection of maps, $f_Z:Z\rightarrow Z'$, $f_Q:Q\rightarrow Q'$, and $f_X:X\rightarrow X'$, such that the chain map commutes with the boundary operators.

We define, for $q\in Q$, that
\be
f_Q(q)=q+\langle q,w \rangle v.
\ee
This map is such that $f_Q(v)=0$, so it ``kills" the corresponding $Z$ stabilizer.

To define $f_Z$, let us first define $Z'$.  In a basis independent fashion, we define $Z'$ to be the subspace of $Z$ containing vectors $z$ such that $\langle \partial_Z z,w \rangle=0$.  We may give it a basis by picking a basis for $Z$ that corresponds to a choice of $Z$-stabilizer generators where all generators commute with $O$, except for one generator corresponding to vector $v$, and then a basis for $Z'$ is chosen 
in the obvious way: the $Z$-stabilizer generators of $C'$ can be chosen to be the $Z$-stabilizer generators of $C$ which commute with $O$.

The boundary operator $\partial'_{Z'}$ is defined in the obvious way from the stabilizers of $C$.  In the basis independent definition of $Z'$, the boundary operator $\partial'_{Z'}$ is the boundary operator $\partial_Z$ on the given subspace.

In the first, basis independent method, we define
\be
\label{fZindep}
f_Z(z)=z+\langle \partial_Z z,w \rangle \partial^{-1} v,
\ee
where $\partial^{-1} v$ is any vector in the pre-image of $v$, chosen arbitrarily (of course, we make this choice one time, and then use the same choice for all $z$).  Note that if there is no redundancy among the stabilizer generators then $\partial^{-1} v$ is 

In the second, basis dependent method, we define $f_Z$ in the obvious way: each basis element of $Z$ corresponding to a generator which commutes with $O$ is mapped to the corresponding basis element of $Z'$, while the basis element which does not commute with $O$ is mapped to zero.
One may verify that \cref{fZindep} indeed defines this map on the basis elements of $Z$ if we choose
$\partial^{-1} v$ to be the basis element .

One may verify the chain map condition that
\be
\partial'_{Z'} f_Z(z)=f_Q(\partial_Z z).
\ee

We may define $X'$ to be the direct sum of $X$ with a $1$-dimensional vector space, corresponding to the one additional $X$-stabilizer in $C'$.  Thus, $X'$ is defined by ordered pairs $(x,b)$ where $x\in X$ and $b\in \{0,1\}$.  We define $\partial'_{Q'}$ in the obvious way:
$$(\partial'_{Q'})^T (x,b)=(\partial_X)^T x + bw,$$
where the superscript $T$ denotes the adjoint,
so $$\partial'_{Q'} q = (\partial_Q q, \langle q,w \rangle).$$

Then define $f_X$ to be the obvious map from $X$ to $X'$:
\be
f_X(x)=(x,0).
\ee
We claim that we also have the chain map condition
\be
\partial'_{Q'} f_Q(q)=f_{X}(\partial_Q q).
\ee 
That is, we claim
$\partial'_{Q'} q + \langle q,w \rangle \partial'_{Q'} v = \langle (\partial_Q q),0 \rangle$.
We have $\partial'_{Q'} v = (0,1)$ since, by assumption, $\langle v,w \rangle=1$ and $\partial_Q v=0$ since $\cC$ is a chain complex (i.e., $v$ corresponds to a $Z$-stabilizer),
and then the claim follows from the definition of $\partial'_{Q'}$.

\section{Toy Model of Vacancies Without Measuring Superstabilizers}
\label{apptoy}
We now consider a toy model of decoding in the presence of vacancies, but \emph{without} measuring superstabilizers (i.e., without measuring logical operators of the code created by the vacancies).  
In this case, anyons can move onto the holes created by the defects without being detected, destroying the error correction properties of the code.  Logical information can be lost in time $\cO(1)$, independently of system size, even if all the holes are size $\cO(1)$, so long as many holes are present.
We are interested in how the logical error probability grows with time if we start from some state with no anyons anywhere and then evolve for some time in the presence of noise without measuring superstabilizers.  We argue that the growth can be superlinear with time in some cases.

While our main interest is in decoding quantum codes in two dimensions (e.g., surface codes, Floquet codes such as the honeycomb code, and so on), we consider here a simpler toy model of decoding \emph{classical} information in a one-dimensional Ising model with vacancies.  To motivate this model, and argue that its behavior should be analogous to that of quantum codes with defects in two dimensions, note that in many stabilizer codes, the defects of the code have certain integer spatial dimensions.  For example, anyons in the two-dimensional toric code are pointlike particles (dimension $=0$), while the three-dimensional toric code has both pointlike (dimension $=0$) and looplike (dimension $=1$) excitations.  The classical Ising model in two-dimensions has one-dimensional domain walls but the one-dimensional Ising model, like the two-dimensional toric code, has pointlike excitations.  Thus, the two-dimensional toric code and the one-dimensional Ising model are in some way similar as the defects in both are pointlike.  Similarly, vacancies in a theory could be pointlike (sites are randomly deleted from a system), linelike (lines are randomly removed), and so on, so vacancies can also be $0,1,2,\ldots$ dimensional.  The questions we consider here arise when the vacancy dimension matches the defect dimension.

We consider the following model of a one-dimensional Ising model with vacancies.  We have $N$ spins, labeled $1,2,\ldots,N$.  Each spin $i$ corresponds to some classical variable $Z_i=\pm 1$.  We assume that the system is initialized in some unknown state, either all spins $+1$ or all spins $-1$, with equal probability for each choice.  Then, the system proceeds through $T$ rounds of errors.  In each round, each spin is flipped independently with some probability $p$.  Then, checks $Z_i Z_{i+1}$ are measured, for each $i \in \{1,2,\ldots,N-1\}\setminus V$, where $V$ is some set of ``vacancies".  That is, we measure checks on every neighboring pair of spins, except for certain vacancies.  Finally, after $T$ rounds, all spins are measured perfectly (i.e., with no error), and we attempt to reconstruct the initial state from those final spin measurements and from the check measurements.

\emph{Vacancies Everywhere---}
As a warmup, we begin with a simple case where $V=\{1,2,\ldots,N-1\}$.  That is, there are vacancies everywhere and no checks are measured.  After $T$ rounds of flips, each spin is flipped from its initial state with probability
\be
\label{Pflip}
P_{flip}=\frac{1-(1-2p)^T}{2}.
\ee
We can regard this as defining some biased coin, which is tails with probability $P_{flip}$ and heads with probability $1-P_{flip}$.  Maximal likelihood decoding in this case is simply majority decoding: we assume that the initial state is where all spins had the value given by the majority after $T$ rounds.
Thus, decoding gives the correct answer if $N$ flips of this biased coin have more than $N/2$ heads and fails if $N$ flips have more than $N/2$ tails.  If $N$ is even and there are exactly $N/2$ heads, then maximum likelihood decoding returns that no decoding is possible: both initial states are equally likely.

Taking the case of odd $N$ for simplicity, the probability of an error is then equal to
\be
P_{err}=\sum_{j=\lfloor N/2 \rfloor+1}^N {N \choose j} P_{flip}^j (1-P_{flip})^{N-j}.
\ee
For $P_{flip}\ll 1$, this is approximated by
$$(P_{flip})^{\lfloor N/2\rfloor+1} {N \choose \lfloor N/2 \rfloor +1} \sim \frac{2^N}{\sqrt{N}}(P_{flip})^{\lfloor N/2\rfloor+1},$$
and for $pT\ll 1$ we have $P_{flip}=pT-cO(T^2)$ so the error probability $P_{err}$ grows \emph{superlinearly} in $T$.  Indeed, it grows as a power $\lfloor N/2 \rfloor +1$.

\emph{Intervals: Generalities---}
We next consider the general case, of arbitrary vacancy locations.  We can describe this by saying that
there are intervals of lengths $\ell_1,\ell_2,\ldots,\ell_k$ for some $k$, i.e. $N=\sum_{i=1}^k \ell_i$ and $V=\{\ell_1,\ell_1+\ell_2,\ell_1+\ell_2+\ell_3,\ldots\}$.
Consider a given interval of some length $\ell$. In a given round, the probability of $m$ errors in that interval is
$$P(m,\ell)=p^m (1-p)^{\ell-m} {\ell \choose m}.$$
Because all checks are present in an interval, the decoder can determine that either $m$ or $\ell-m$ errors occurred; indeed, it knows that either some specific pattern of $m$ errors occurred or the ``complementary" pattern of $\ell-m$ errors occurred.
The decoder can then correct the errors that occurred, up to possibly an overall flip of all spins in the interval; if the decoder chooses to correct the minimal number of errors (i.e., for $m<\ell/2$, it assumes $m$ errors rather than $\ell-m$ occurred), then this can be described by flipping a biased coin: with
probability
$P(\ell-m,\ell)/(P(m,\ell)+P(\ell-m,\ell))$ the value of all spins in the interval gets flipped.

Thus, we can describe this by the following effective model (and we emphasize that a maximal likelihood decoding of this effective model will give a maximum likelihood decoding of the original model, so that no meaningful information is lost when going to the effective model).  
We describe the state of each interval by a single \emph{effective spin}, $\pm 1$.
We then draw some integer $m$ at random, $0\leq m\leq \ell/2$, from the following distribution.
For $m<\ell/2$, the probability of having that given $m$ is $P(m,\ell)+P(\ell-m,\ell)$,
while for $m=\ell/2$, which is possible if $\ell$ is even, the probability of having that given
$m$ is $P(m,\ell)$.
The decoder knows the value of this $m$.  Then, for $m<\ell/2$, we toss a biased coin and flip the
effective spin in that interval with probability
$P(\ell-m,\ell)/(P(m,\ell)+P(\ell-m,\ell))$.  If $m=\ell/2$, we flip the effective spin with probability $1/2$.
We emphasize, there is only a single effective spin for each interval.

Thus, in this effective model, each interval behaves independently over the $T$ rounds.  Each interval is subject to a sequence of biased coin tosses, with the bias known to the decoder, flipping the effective spin in the interval if the coin is tails.

Let
$P_{flip}(i,t)$ denote the probability of flipping the effective spin in the $i$-th interval on the $t$-th round.
Remark: in the previous case where we had vacancies everywhere, all $P_{flip}(i,t)$ are equal to each other (indeed, they equal $p$), but now we allow them to vary from round to round and from interval to interval.
Thus, the effect of $T$ rounds of flips on a given effective spin is that it is flipped, from its initial state at round $0$, with probability
\be
\label{Pflipgen}
P_{flip}(i)=\frac{1-\prod_{t=1}^T\Bigl(1-2P_{flip}(i,t)\Bigr)}{2}.
\ee
This reduces to \cref{Pflip} if $P_{flip}(i,t)=p$ for all $i,t$.
Again, these probabilities $P_{flip}(i)$ are known to the decoder.

Finally, we apply maximum likelihood decoding after $T$ rounds.  The decoder can determine that either some given pattern of errors on the \emph{effective} spins occurred (i.e., some specific set of effective spins are flipped) or that the complementary pattern of errors occured (i.e., that the complement of that set of effective spins is flipped).
Maximum likelihood decoding gives an error if some pattern occurs but the complementary pattern is more likely.  If the complementary pattern is equally likely, then maximum likelihood decoding knows that both initial states were equally likely.

\emph{Two Intervals---}
The above analysis of intervals gives some general results reducing to an effective model.  However, this effective model still must be analyzed: to determine asymptotics, we need to determine which sequences of biased coin flips are most likely.  We work now on the case of two intervals, of lengths $\ell_1,\ell_2$.

We assume without loss of generality that $\ell_1\leq \ell_2$.

There are two ways in which an error can occur.  One way is when the effective spin in both intervals is flipped after $T$ time steps.  The second way is when the effective spin in one interval is flipped while the other is not flipped, but the decoder still makes an error as it determines that it is more likely that the other effective spin flips.  Remark: we say ``is flipped" rather than ``flips" to emphasize that the important thing is whether there is a total of an odd number of flips after $T$ rounds.

Let us consider the first way first.  \cref{Pflipgen} gives the probablity of this occuring for each choice of $P_{flip}(i,t)$.  However, since \cref{Pflipgen} is linear in each $P_{flip}(i,t)$, we can compute the probability that the effective spin is flipped for random choice of $P_{flip}(i,t)$ by replacing $P_{flip}(i,t)$ with its average.
Thus, the probability that the effective spin in an interval of length $\ell$ is flipped is
given by
$$\frac{1-(1-2P_{eff})^T}{2},$$ where for odd $\ell$ we have
$$P_{eff}=\sum_{0<=m<\ell/2} 
\Bigl( P(m,\ell)+P(\ell-m,\ell) \Bigr)\Bigl(
\frac{P(\ell-m,\ell)}{(P(m,\ell)+P(\ell-m,\ell))}\Bigr)
=
\sum_{0<=m<\ell/2} 
P(\ell-m,\ell).$$

From now on, we consider the limiting case $p\ll 1$, where this is dominated by the contribution with $m=\lfloor \ell/2\rfloor$, so
in this case
$$P_{eff} \approx  C p^{\lfloor \ell/2 \rfloor+1} 2^{\ell} \ell^{-1/2},
$$
where the factor of $C 2^{\ell} \ell^{-1/2}$ is an approximation to ${\ell \choose \lfloor \ell/2 \rfloor}$
and $C=\sqrt{2}/\pi$ from Stirling's formula.

Let $$P_{eff}^1=C p^{\lfloor \ell_1/2 \rfloor+1} 2^{\ell_1} \ell_1^{-1/2},$$
and
$$P_{2,eff}=C p^{\lfloor \ell_2/2 \rfloor+1} 2^{\ell_2} \ell_2^{-1/2}.$$

Since $\ell_1\leq \ell_2$, $P_{2,eff}\leq P_{1,eff}$.
We will consider three different asymptotic time regimes below: first, $T P_{1,eff}\ll 1$; second, $T P_{2,eff} \ll 1 \ll T P_{1,eff}$; and finally, $1\ll T P_{2,eff}$.

The probability that both effective spins are flipped is
\be
\label{Pboth}
P_{both}\approx
\Bigl(\frac{1-(1-2 P_{1,eff})^T}{2}\Bigr)
\Bigl(\frac{1-(1-2 P_{2,eff})^T}{2}\Bigr).
\ee

Each term in the product in \cref{Pboth} asymptotes as $T\rightarrow \infty$ at $1/2$ so the product asymptotes at $1/4$.
The first term displays a linear behavior for $P_{1,eff} T\ll 1$, and is roughly constant for $P_{1,eff} T \gg 1$ and similarly for the second term.

In the regime $P_{1,eff} T\ll 1$ we see a quadratic growth in $P_{both}$:
$$P_{both} \sim T^2 P_{1,eff} P_{2,eff}.$$

In the regime $P_{1,eff} T \gg 1$ but $P_{2,eff} T\ll 1$, we see a linear growth in $P_{both}$:
$P_{both} \approx \frac{1}{2} T P_{2,eff}$.

Finally, in the regime $P_{2,eff} T \gg 1$, $P_{both} \approx 1/4$.

Now, let us consider the second way, where one effective spin is flipped and the other is not, but the decoder still decodes incorrectly.

First consider the regime $P_{1,eff} T\ll 1$.  Suppose that a single coin toss at some time in one of the intervals is tails, and all other tosses are heads.  
To leading order in $P_{1,eff} T$, it suffices to consider this possibility.

Suppose for  example, a coin toss in the first interval is tails (the case where a toss in the second interval is tails is similar to this case), and occurs at some time
when we have some $m_1 \leq \ell_1/2$ errors in the first interval at some time.
If at any time we have
$m_2 \leq \ell_2/2$ errors in the second interval with $\ell_1/2-m_1>\ell_2/2-m_2$ then the decoder will make an error because in this case it is \emph{more likely that the first coin toss is heads and the second is tails, than visa-versa}.

The probability that this occurs (i.e., that we have some given $m_1$ at some time, that that coin toss is tails, and that we have such an $m_2$ at some, possibly different, time) is at the same order in the physical error error probability as in the case above where we made an error in each interval.  That is, it is at order $p^{\lfloor \ell_1/2 \rfloor + \lfloor \ell_2/2\rfloor + 2}$.
Further, for $m_1 \approx \lfloor \ell_1/2\rfloor$ and $m_2 \approx \lfloor \ell_2/2\rfloor$, the probability of an error here is similar to that in the case above: the probability that such $m_1,m_2$ occur gives roughly the same quadratic growth in time as both events ($m_1\approx \lfloor \ell_1/2 \rfloor$ and $m_2 \approx \lfloor \ell_2/2 \rfloor$) are unlikely and there are $T$ different times that each could occur at.

However, we could also have a case with $\ell_1=\ell_2$ and have an event where at some time with $m_1=0$ the coin toss in the first interval is tails.  While this is event is rare (having small $m_1$ is not rare, but having the toss be tails is rare), the decoder can make a mistake without any rare event in the second interval occuring (as it can make a decoding mistake even if $m_2$ is always small in this case).  So, this kind of contribution gives rise to only a linear growth in $T$.
However, as $m_1$ becomes smaller than $\ell_1/2$ or $m_2$ becomes smaller than $\ell_2/2$,
the probability of these events is becomes \emph{less likely} because the factor of
${\ell_1 \choose m_1} {\ell_2 \choose m_2}$ is smaller.
So, the dominant contribution is quadratic in $T$, as it was when we considered the first way of making an error.

In the regime $P_{2,eff} \ll 1 \ll P_{1,eff}$, it is likely that several times we had $m=\lfloor \ell_1/2 \rfloor$ in the first interval.  Indeed, the expected number of times that this occurs is proportional to $P_{1,eff} T$.  In this case, a maximal likelihood decoder knows that the first effective spin likely flips several times\footnote{There is an exponentially small (in $P_{1,eff} T$) probability that we never see $m=\lfloor \ell_1/2 \rfloor$; this possibility does not contribute significantly to the decoding error probability.}.  
Indeed, the first effective spin is flipped with probability close to $1/2$.
So, a maximal likelihood decoder will typically use the second effective spin to determine the correct decoding, largely ignoring the first.  The probability that the second effective spin is 
flipped is approximately $T P_{2,eff}$ in this regime.  So, the probability of an error occurring in this way is roughly $(1/2) T P_{2,eff}$.

Thus, counting both ways of making an error in this regime (in the first way, both effective spins are flipped, while in the second way the dominant contribution is that the first effective spin is not flipped and the second effective spin is flipped), we claim that for 
$P_{1,eff} T \gg 1$ but $P_{2,eff} T\ll 1$, there is a linear growth in the probability of an error in decoding so that it is roughly
$(1/2) T P_{2,eff}$.

Similarly, in the regime $P_{2,eff} T \gg 1$, the probability of an error in decoding is roughly $1/2$ when we consider both ways of making an error.

So, there is a crossover from quadratic, to linear, to constant error probability as $T$ increases.

\bibliography{sc-ref}
\end{document}